\documentstyle[preprint,prd,eqsecnum,aps]{revtex}
\tighten
\begin{document}
\draft

\title {A vacuum-like configuration in General Relativity\\
as a manifestation of a Lorentz-invariant mode\\ of
five-dimensional gravity}
\author{Valentin D. Gladush\footnote{E-mail: vgladush@ff.dsu.dp.ua}}
\address{Physics Department, Dnepropetrovsk National University, \\
Naukova Str. 13, Dnepropetrovsk 49050, Ukraine}
\date{\today}
\maketitle

\begin{abstract}
A Lorentz-invariant cosmological model is constructed within the
framework of five-dimensional gravity. The five-dimensional theorem
which is analogical to the generalized Birkhoff theorem is proved,
that corresponds to the Kaluza's ``cylinder condition''. The
five-dimensional vacuum Einstein equations have an integral of motion
corresponding to this symmetry, the integral of motion is similar to
the mass function in general relativity (GR). Space closure with
respect to the extra dimensionality follows from the requirement of
the absence of a conical singularity. Thus, the Kaluza-Klein (KK)
model is realized dynamically as a Lorentz-invariant mode of
five-dimensional general relativity. After the dimensional reduction
and conformal map\-ping the model is reduced to the GR configuration.
It contains a scalar field with a vanishing conformally invariant
energy-momentum tensor on the flat space-time background. This zero
mode can be interpreted as a vacuum configuration in GR. As a result
the vacuum-like configuration in GR can be consi\-dered as a
manifestation of the Lorentz-invariant empty five-dimensional space.
\end{abstract}

 \narrowtext

\section{INTRODUCTION}

The data of the observation astronomy and cosmology indicate the
existence of new forms of matter in the Universe such as ``dark
matter'' and ``dark energy'' (for a review see \cite{turner}). One of
the candidates for the dark energy role is vacuum. It is associated
with a non-vanishing cosmological constant (see \cite{weinberg},
\cite{chernin}, \cite{sahni}, and references therein). The vacuum is
defined as a Lorentz invariant state of physical medium in space-time
$V^{(4)} $\cite{zeldovich}, \cite{gliner1}, \cite{gliner2}. This
concept has become a basis for the popular model of inflation in the
early Universe \cite{guth}, \cite{linde90}, \cite{brandenberger}.
Here the vacuum configuration is a cosmological model with the metric
admitting the Lorentz group as the isometry group. The gravitational
field is generated by the Lorentz-invariant ``vacuum-like''
energy-momentum tensor
\begin{equation}\label{TEi}
  T^{\mu}_{\nu}=\frac{\Lambda}{8\pi G}\, \delta^{\mu}_{\nu}\,
\end{equation}
where $G$ is the Newtonian gravitational constant, $\Lambda $ is the
cosmological constant. The vacuum density is determined by this
constant: $\rho_{v}=\Lambda/8\pi G $ ($c=1 $).

Let the Lorentz group $L=O(1,3)$ act on the coordinates $x^{\mu}$ $(\mu\,,\nu =
0, 1, 2, 3)$ of space-time $V^{(4)}$ in the following way
\begin{equation}\label{loren}
  \tilde{x}^{\mu}=L^{\tilde{\mu}}_{\nu}x^{\nu}\,.
\end{equation}
Moreover, $\eta_{\tilde{\mu}\tilde{\nu}}  =
  L_{\tilde{\mu}}^{\alpha}L_{\tilde{\nu}}^{\beta}\eta_{\alpha\beta}\,,
  \ L_{\tilde{\mu}}^{\alpha}L^{\tilde{\mu}}_{\beta}
  =\delta^{\alpha}_{\beta}\,,\
  L_{\tilde{\nu}}^{\alpha}L^{\tilde{\mu}}_{\alpha}
  =\delta^{\tilde{\mu}}_{\tilde{\nu}}$
where $\eta_{\mu\nu}$, $\eta_{\tilde{\mu}\tilde{\nu}}$ are Minkowski
metrics. The invariance condition $V^{(4)}$ with respect to the
Lorentz group leads to the conformally flat space with the metric
\begin{equation}\label{4metrkonf}
  ^{(4)}ds^2=g_{\mu\nu}dx^{\mu}dx^{\nu}
  =\psi^2(\zeta)\eta_{\mu\nu}dx^{\mu}dx^{\nu}
\end{equation}
where
\begin{equation}\label{zetaxx}
  \zeta=\eta_{\mu\nu}x^{\mu}x^{\nu}\,.
\end{equation}

The Weyl tensor vanishes for the metric (\ref{4metrkonf}). This leads to the
curvature tensor ${^{(4)}R}_{\mu\,\nu\,\rho\,\sigma}= {^{(4)}R}_{\rho[\mu}\,
g_{\nu]\sigma}-{^{(4)}R}_{\sigma[\mu}\, g_{\nu]\rho} -(1/3){^{(4)}R}\,
g_{\rho[\mu}\, g_{\nu]\sigma}$ where the square brackets mean an alternation,
i.e. $A_{[\mu\nu]}=(A_{\mu\nu}-A_{\nu\mu})/2$, $^{(4)}\!R_{\mu\nu}$ is the
Ricci tensor, $^{(4)} \!R$ is the scalar curvature. From the Einstein equations
\begin{equation}\label{4d-eqg0}
  ~^{(4)}G_{\mu\nu}=\,^{(4)}\!R_{\mu\nu}
  -(1/2) \,^{(4)} \!R\ g_{\mu\nu}=\Lambda g_{\mu\nu}\,
\end{equation}
it follows that $^{(4)}\!R_{\mu\nu} = - \Lambda g_{\mu\nu}$ è $^{(4)}
\!R=-4\Lambda \,$. Therefore ${^{(4)}R}_{\mu\,\nu\,\rho\,\sigma} =
K(g_{\mu\rho}\, g_{\nu\sigma}-g_{\mu\sigma}\, g_{\nu\rho})$ where
$K=-1/a^2=-\Lambda/3$ and we come to the de Sitter space-time (see
\cite{hawking} for details).

In the generalized stereographic coordinates \cite{gladush7},
\cite{karl} the metric of constant curvature spaces can be written
down as follows \cite{petrov}
\begin{equation}\label{metrkonfor}
  ^{(4)}ds^2 = g_{\mu\nu}dx^{\mu}dx^{\nu}
  = {\left(1-{\Lambda\zeta}/{12}\,\right)^{-2}}\
  {\eta_{\mu\nu}dx^{\mu}dx^{\nu}}\,.
\end{equation}
The one-parameter family of vacuum-like configurations with the
metric (\ref{metrkonfor}) is labeled by the parameter $\Lambda $. The
last defines the vacuum density $\rho_{v}=\Lambda/8\pi G$. At present
the vacuum density value $\rho_{v}$ (or the constant $\Lambda $) is
determined only from observations \cite{chernin}.

It turns out that it is possible to construct a nontrivial ``vacuum''
configuration without the cosmological constant. Indeed,  we consider
a scalar field  on the flat space-time background. Let the scalar
field is a Lorentz-invariant solution of the wave equation and has a
vanishing conformally invariant energy-momentum tensor. This
configuration appears as a zero mode of the GR equations. In the
present paper the interpretation of classical vacuum in GR as a
Lorentz-invariant zero mode of the GR equations for the coupled
gravitational and scalar fields with conformal connection is proposed
with using this vacuum-like configuration.

The model admits a geometrical formulation within the framework of
five-dimensional gravity \cite{wes}. It is shown that the
Lorentz-invariant Kaluza-Klein space can be considered as a
Lorentz-invariant mode of the equations of five-dimensional gravity
(without external sources) where the cylindricity and closedness
conditions are realized
dynamically. After the dimensional reduction and corresponding
conformal transformation this model is reduced to the above
four-dimensional vacuum configuration with a scalar field on the flat
space-time background. Therefore the vacuum in GR can be interpreted
as a manifestation of the Lorentz-invariant mode of five-dimensional
gravity or that of the KK theory.

The plan of the paper is following. In Sec. \ref{model} we introduce the
necessary relations used in five-dimensional gravity for the vacuum
Lorentz-invariant five-dimensional spaces. In Sec. \ref{birkhoff} we establish
the five-dimensional version of the generalized Birkhoff theorem and explicitly
construct an extra Killing vector $\vec {\xi}$.

In Sec. \ref{D5masfunct} we find the `first integral'' of the
five-dimensional vacuum Einstein equations which leads to the
conservation law $G=L^2 (\nabla L)^2 -\epsilon L^2= \mbox{const}$. In
Sec. \ref{5Dmodel} we obtain the metric for conformally flat
five-dimensional Lorentz-invariant cosmological model and ascertain
the physical meaning of $G$. After the dimensional reduction and
corresponding conformal transformation we reduce this model to the
four-dimensional vacuum-like Lorentz-invariant configuration for the
gravitational and conformal scalar fields with conformal connection.
In Sec. \ref{geometry} we consider some geometrical properties of the
model and the regularity condition. In Sec. \ref{geodesic} we reduce
the five-dimensional action for a geodesic line to the
four-dimensional action for a particle with a variable mass in the
flat space-time.

In Appendix $A $ the generalized stereographic coordinates are
introduced. In Appendix $B $ the curvature tensor, the Ricci tensor,
the scalar curvature, and also the quadratic curvature invariant for
the obtained metric are calculated. In Appendix $C $ the proof of the
five-dimensional version of the Birkhoff theorem is completed.

Here $G$ is the gravitational constant, the velocity of light $c=1$. The metric
tensor $g_{\mu\nu}\, (\mu,\nu=0,1,2,3)$ has a signature (+~-~-~-).

\section{The basic equations of Lorentz-invariant\\ five-dimensional
cosmological model}\label{model}

In the framework of the five-dimensional general relativity \cite{wes} we consider the
five-dimensional pseudo-Riemannian space $V^{(5)}$ with the metric $
^{(5)}\!g_{AB}$ ($A,~B=0$, 1, 2, 3, 4) (the signature is $(+ - - - -)$) which
in general depends on all the coordinates $x^A $. This metric satisfies the
five-dimensional Einstein equations in the vacuum
\begin{equation}\label{2.1}
  ~^{(5)}G^{A}_{B}=\,^{(5)}\!R^{A}_{B}
  -\frac{1}{2}\ \,^{(5)} \!R\ \delta^{A}_{B}=0\,.
\end{equation}
These equations follow from the five-dimensional Hilbert variational
principle for the action
\begin{equation}\label{act}
I^{(5)}=-\frac{1}{16\pi \hat{G}}\int \sqrt{|^{(5)}g|}~^{(5)}R\ d^5x
\end{equation}
where $\hat{G}$ is the ``five-dimensional gravitational constant'',
$^{(5)}g = \det(^{(5)}g_{AB})$. We further\-more suppose that
$x^{\mu}$ $(\mu\,,\nu = 0, 1, 2, 3)$ are space-time coordinates and
$x^{4}\equiv Z$ is the extra fifth coordinate.

Now we consider the five-dimensional Lorentz-invariant cosmological
model. The Lorentz group $L$ acts on space $V^{(5)}$ by the rule
(\ref{loren}). The invariance condition of $V^{(5)}$ with respect to
the Lorentz group leads to the space admitting the families of
conformally flat space-time local subspaces. Therefore, the
five-dimensional metric can be written via generalizing the metric
(\ref{metrkonfor}) of space-time $V^{(4)}$. It is easy to see that
the desired maximum general Lorentz-invariant metric $V^{(5)}$ can be
written in the following form
\begin{equation}\label{metru}
^{(5)}ds^2 = \psi^{2}(Z,\zeta)\eta_{\mu\nu}dx^{\mu}dx^{\nu}
    -\tilde{V}^{2}(Z,\zeta)\left(d Z+N(Z,\zeta)x_{\mu}dx^{\mu}\right)^2
\end{equation}
where $\psi, \,\tilde{V},\,N$ are some functions,
$x_{\mu}=\eta_{\mu\nu}x^{\nu}$. According to the Frobenius theorem
\cite{kram}, we have $d Z+N(Z,\zeta)x_{\mu}dx^{\mu}
  =d Z+\tilde{N}(Z, u_{\epsilon}) d u_{\epsilon}
  =\beta d z $
where $\tilde{N}(Z, u_{\epsilon})=u_{\epsilon}N(Z,\epsilon u_{\epsilon}^2) $,
$z$ and $\beta$ are some functions of variables $Z$ and $\zeta$. Using a
function $z=z(Z,\zeta) $ as a new fifth coordinate, we can write the metric
(\ref{metru}) in the for
\begin{equation}\label{metr}
   ^{(5)}ds^2 = {^{(5)}g}_{AB}dx^{A}dx^{B}
   =\psi^{2}(z,\zeta)\eta_{\mu\nu}dx^{\mu}dx^{\nu}
    -V^{2}(z,\zeta)dz^2
\end{equation}
where $V^2=\beta \tilde{V}^2$. In \ref{appen1} it will be shown that the
relation
\begin{equation}\label{flate-conf}
^{(4)}ds^2_{0}=\eta_{\mu\nu}dx^{\mu}dx^{\nu} =\epsilon\,
du^2_{\epsilon} - u^2_{\epsilon}\,{^{(3)}\!d\Omega}^2_{\epsilon}\,
\end{equation}
takes place. Here
\begin{equation}\label{eu2}
  u_{\epsilon}=\sqrt{\epsilon\eta_{\mu\nu}x^{\mu}x^{\nu}}
  =\sqrt{\epsilon\zeta}\, \qquad
    (\epsilon ={\zeta}/{|\zeta|})\,
\end{equation}
is the ``radial'' variable which labels the hyperboloids
$\epsilon\eta_{\mu\nu}x^{\mu}x^{\nu}=u^2_{\epsilon}$,
\begin{equation}\label{domegapm}
   ^{(3)}\!d\Omega^{2}_{\epsilon}= h_{ij}dy^i dy^j
   = {\left(1-\epsilon{S_{\epsilon}^2}/{4}\, \right)^{-2}}
   ({\epsilon (dy^1)^2 +(dy^2)^2 +(dy^3)^2} )
\end{equation}
is the ``angular'' part of metric (\ref{flate-conf}) in the
stereographic coordinates $y^i$ \cite{gladush7}, \cite{karl}, $
S_{\epsilon}^2=\epsilon(y^1)^2 +(y^2)^2 +(y^3)^2 $,  (i,j=1,2,3). The
metric (\ref{metr}) can be written in the form
\begin{equation}\label{pmmetr}
^{(5)}ds^2 = \psi^{2}(z,\zeta)(\epsilon\, du^2_{\epsilon}
    - u^2_{\epsilon} \,{^{(3)}\!d\Omega}^2_{\epsilon})
    -V^{2}(z,\zeta)dz^2
\end{equation}
with taking into account (\ref{flate-conf}).

One can find the symmetry of the model by solving the five-dimensional Killing
equations
\begin{equation}\label{eqtKl}
 {\cal L}_{\vec{\xi}}\, ^{(5)}\!g_{AB}=
  {^{(5)}\!g_{AB}}_{,\,C}\xi^{C}+ {^{(5)}\!g_{AC}}\,\xi^{C}_{,\,B}
  + {^{(5)}\!g_{CB}}\,\xi^{C}_{,\,A}=\xi_{A;\,B}+\xi_{B;\,A}=0
\end{equation}
where ${\cal L}_{\vec{\xi}}\,$ is a Lie derivative with respect to
the Killing vector $\vec{\xi}=\xi^{A}\partial _{A}=\xi^{z}\partial
_{z}+\xi^{\mu}\partial _{\mu}$. Here ``$\, _{,\,A}$''
$\equiv\partial_{A} $ $\equiv\partial/\partial x^{A} $ and ``$\,
_{,\,z}$'' $\equiv\partial_{z} \equiv\partial/\partial z $ denote the
partial derivatives with respect to the coordinates $x^A $ and $z $,
correspondingly. For the metric (\ref{metr}) these equations take the
form
\begin{eqnarray}
    (V\xi^{z})_{,\,z} & + & V_{,\,\mu}\,\xi^{\mu}=0\,, \label{eqkl1}\\
    V^2\xi^{z}_{,\,\mu} & - & \psi^2\xi_{\mu,\,z}=0\,, \label{eqkl2} \\
   2\eta_{\mu\nu}(\psi_{,\,z}\xi^{z} & + & \psi_{,\,\alpha}\xi^{\alpha})
     +\psi(\xi_{\mu,\,\nu}+\xi_{\nu,\,\mu})=0  \label{eqkl3}
\end{eqnarray}
where $\xi_{\mu}=\eta_{\mu\nu}\xi^{\nu}$. The general nontrivial
solution of these equations
   $ \vec{\xi}_{L} = x_{\mu}\omega^{\mu\nu}\partial_{\nu}\,\
      (x_{\mu}=\eta_{\mu\nu}x^{\nu})$
for arbitrary $\psi(Z,\zeta)$, $V(Z,\zeta)$ determines a general member of the
Lie algebra of the Lorentz group. Here $\omega^{\mu\nu}=-\omega^{\nu\mu}$ are
parameters fixing a member of this algebra.

For the metric (\ref{metr}) the Einstein tensor have the form (see
\ref{appen3})
\begin{eqnarray}
 & ^{(5)}G^{\mu}_{\nu} ={^{(5)}G}^{u}_{u}u_{\nu}^{\epsilon}u^{\mu}_{\epsilon}
  +{^{(5)}G}^{h}_{h}h_{\nu}^{\mu}\,,\label{5G^mn}\\
 & ^{(5)}G^{z}_{\mu}= {^{(5)}\!R^{z}_{\mu}}
  ={^{(5)}G}^{z}_{u}u^{\epsilon}_{\mu}\,,
  \quad ^{(5)}G_{z}^{\mu}
  =\epsilon\,{^{(5)}\!G}_{z}^{u}u_{\epsilon}^{\mu}\,,\label{5G^5m}\\
 & ^{(5)}G^{z}_{z}= {6}{\psi^{-3}}\left(
  {2\zeta\psi_{\zeta\zeta}}
  +{4\psi_{\zeta}}-{\psi V^{-2}\psi^2_{z}} \right) \label{5G^55}
\end{eqnarray}
where
\begin{eqnarray}
  {^{(5)}G}^{z}_{u} & = & -\frac{\epsilon\psi^2}{V^2}{^{(5)}G}^{u}_{z}
      =  \epsilon{^{(5)}G}^{z}_{\mu}u^{\mu}_{\epsilon} = 
   \frac{6u_{\epsilon}}{\psi V^2}\left(\psi_{z\zeta}-
  \frac{\psi_{z}\psi_{\zeta}}{\psi} \right.
  - \left.\frac{\psi_{z}V_{\zeta}}{ V}\right)
  =\frac{6u_{\epsilon}}{V}
  \left(\frac{\psi_{z}}{\psi V}\right)_{\zeta}\,,\label{5G^zu}\\
  {^{(5)}G}^{u}_{u}
 & = & {^{(5)}G}^{\nu}_{\mu}u^{\mu}_{\epsilon}u_{\nu}^{\epsilon} =
  \frac{3\epsilon}{\psi^2}\left(-\frac{\psi\psi_{zz}}{V^2}-\frac{\psi^2_{z}}{V^2}
   +\frac{\psi\psi_{z}V_{z}}{V^3} \right. 
  +  \left. \frac{4\zeta\psi^{2}_{\zeta}}{\psi^2}
   +\frac{4\psi_{\zeta}}{\psi}+\frac{4\zeta\psi_{\zeta}V_{\zeta}}{\psi V}
   +\frac{2V_{\zeta}}{V} \right)\,,\label{5G^uu}
\end{eqnarray}
\begin{eqnarray}
  ^{(5)}G^{h}_{h}  =
  \frac{1}{\psi^2}\left(
  \frac{3\psi\psi_{zz}}{V^2}+\frac{3\psi^2_{z}}{V^2}
    - \frac{3\psi\psi_{z}V_{z}}{V^3}
    -  \frac{4\zeta V_{\zeta\zeta}}{V}\right.
  & - &  \left.\frac{6V_{\zeta}}{V}- \frac{8\zeta\psi_{\zeta\zeta}}{\psi}\right.
   \nonumber\\
   & + & \left.\frac{4\zeta\psi^{2}_{\zeta}}{\psi^2}
  -\frac{12\psi_{\zeta}}{\psi}
  -\frac{4\zeta\psi_{\zeta}V_{\zeta}}{\psi V}\right)\,.\label{5G^hh}
\end{eqnarray}
Here $u^{\mu}_{\epsilon} ={x^{\mu}}/{u_{\epsilon}}\ $
$(u_{\mu}^{\epsilon}=\eta_{\mu\nu}u_{\epsilon}^{\nu})$ and
``$_{z}$''$\equiv\partial/\partial{z}$,
``$_{zz}$''$\equiv\partial^2/\partial{z}^2$,
``$_{\zeta}$''$\equiv\partial/\partial{\zeta}$,
``$_{\zeta\zeta}$''$\equiv\partial^2/\partial{\zeta}^2$. Operations with
indexes for quantities $A^{u }$ and $A_{u}$ are given by the formula
(\ref{u^-uac}).

\section{Five-dimensional version of the generalized\\ Birkhoff theorem as a
cylinder condition} \label {birkhoff}

If the metric (\ref{metr}) satisfies the vacuum equations of five
dimensional general relativity, the symmetry of the model turns out
to be higher than the initial one. In this case the metric
(\ref{metr}) admits extra one-parameter isometry group with the
Killing vector $\vec{\xi}=\xi^{A}\partial _{A}=\xi^{z}\partial
_{z}+\xi^{\mu}\partial _{\mu}$. This group should commute with the
Lorentz group, therefore we have
\begin{equation}\label{comutat}
 [\vec{\xi},\vec{\xi}_{L}]=0
\end{equation}
whence, using the formula $\vec{\xi}_{L} =
x_{\mu}\omega^{\mu\nu}\partial_{\nu}$, we obtain the following defining
equations
\begin{equation}\label{kill-equ}
   x_{\mu}\omega^{\mu\nu}\xi^{z}_{,\,\nu}=0\,,\qquad
   \xi_{\mu}\omega^{\mu\nu}
   - x_{\mu}\omega^{\mu\rho}\xi^{\nu}_{,\,\rho}=0\,.
\end{equation}
The solution of these equations has the form
\begin{equation}\label{solkilequ}
  \xi^{z}=h(z,\zeta)\,,\qquad \xi^{\nu}=f(z,\zeta)x^{\mu}
\end{equation}
where $h=h(z,\zeta)$, $f=f(z,\zeta)$ are some functions. Therefore the Killing
vector can be written as
\begin{equation}\label{solcomut}
  \vec{\xi}=h\,\partial_{z}+fx^{\mu}\, \partial_{\mu}\,.
\end{equation}
After substitution (\ref{solkilequ}), the Killing equations
(\ref{eqkl1}) -- (\ref{eqkl3}) are converted as:
\begin{eqnarray}
    (V h)_{,\,z} & + & fV_{,\,\mu}\,x^{\mu}=0\,, \label{eqkil1L}\\
    V^2h_{,\,\mu} & - & \psi^2f_{,\,z} x_{\mu}=0\,, \label{eqkil2L} \\
   2\eta_{\mu\nu}(h\,\psi_{,\,z} & + & f\psi_{,\,\alpha}x^{\alpha}+f\psi)
     +\psi f_{,\,\nu}x_{\mu}+\psi f_{,\,\mu}x_{\nu}=0\,. \label{eqkil3L}
\end{eqnarray}

For arbitrary functions $\Phi=\Phi(z,\zeta)$ the formulae take place
\begin{equation}\label{deriv}
  \Phi_{,\,\mu}=2 \Phi_{,\,\zeta}x_{\mu}
  =\epsilon \Phi_{,\,u_{\epsilon}}u^{\epsilon}_{\mu}\,,\qquad
  x^{\mu}\Phi_{,\,\mu}=2\zeta \Phi_{,\,\zeta}
  =u_{\epsilon}\Phi_{,\,u_{\epsilon}}\,
  \quad (\zeta=\epsilon u_{\epsilon}^2=\eta_{\mu\nu}x^{\mu}x^{\nu})\,.
\end{equation}
Therefore the relations (\ref{eqkil1L}) -- (\ref{eqkil3L}) can be
rewritten as follows
\begin{eqnarray}
(V h)_{,\,z} & + & 2 f\zeta V_{,\,\zeta}=0\,,\label{eqkil1L2}\\
    2 V^2h_{,\,\zeta} & - & \psi^2 f_{,\,z}=0\,, \label{eqkil2L2} \\
   \eta_{\mu\nu}(h\,\psi_{,\,z} & + & f(\psi+2\zeta\psi_{,\,\zeta}))
     +2\psi f_{,\,\zeta}x_{\nu}x_{\mu}=0\,. \label{eqkil3L2}
\end{eqnarray}
Using the formulae (\ref{uhmn}) and (\ref{nm}), we obtain from the last
equation
\begin{eqnarray}\label{eqkil3La}
 \left(h\,\psi_{,\,z}+f(u_{\epsilon}\psi)_{,\, u_{\epsilon}})\right)
 (\epsilon u^{\epsilon}_{\mu}u^{\epsilon}_{\nu}-h_{\mu\nu})
 +2\psi u^{2}_{\epsilon}f_{\zeta}
u^{\epsilon}_{\mu}u^{\epsilon}_{\nu}=0\,.
\end{eqnarray}
Hence it follows that
\begin{eqnarray}
    && h\,\psi_{,\,z}+2\epsilon fL_{,\, \zeta}\sqrt{\epsilon\zeta}
    +2\zeta\psi f_{\zeta}=0\,, \label{eqkil3L1x}
\\ && h\,\psi_{,\,z}+2\epsilon fL_{,\, \zeta}\sqrt{\epsilon\zeta}
=0  \label{eqkil3L2x}
\end{eqnarray}
where
\begin{equation}\label{ll}
  L =u_{\epsilon}\psi=\psi\sqrt{\epsilon\zeta}\,.
\end{equation}
The equations (\ref{eqkil3L1x}), (\ref{eqkil3L2x}) are equivalent to
the relations
\begin{equation}\label{eqkil3L2a}
     h\,L_{,\,z}+2f\zeta L_{,\, \zeta} =0\,, \qquad
 f_{,\, \zeta}=0\,.
\end{equation}

Note that the continuity equation
\begin{equation}\label{1divkill}
    \left(\sqrt{^{(5)}g}\,\xi^{A}\right)_{,\,A}=0
\end{equation}
follows from the Killing equations (\ref{eqtKl}). Taking into account the
metric (\ref{metr}) and Killing vector (\ref{solcomut}), we obtain from here
\begin{equation}\label{2divkill1}
  (hV\zeta\psi^4)_{,\,z}
  +2(fV\zeta^2\psi^4)_{,\,\zeta}=0\,.
\end{equation}
This equation will be satisfied identically if we assume
\begin{equation}\label{soleqkin}
  h = \frac{2F_{,\,\zeta}}{ V\zeta\psi^4}\,,\qquad
  f =-\frac{F_{,\,z}}{V\zeta^2\psi^4}
\end{equation}
where $F$ is some function of $z$ and  $\zeta$. Substituting the
expressions for $h$ and $f$ into Eq. (\ref{eqkil3L2a}), we find
$F_{,\,u_{\epsilon}}\,L_{,\,z}-F_{,\,z}L_{,\, u _ {\epsilon}} =0 $.
Hence it follows that $F=F (L) $. As a result, for the Killing vector
(\ref{solcomut}) we have
 $ \vec{\xi}=V^{-1}L^{-4}{F_{,\,L}}
  \left(2\zeta L_{,\,\zeta}\,\partial_{z}
  -L_{,\,z}\,x^{\mu}\,\partial_{\mu}\right)$.

The remaining  Killing equations (\ref{eqkil1L2}), (\ref{eqkil2L2}) and the
last equation in (\ref{eqkil3L2a}) lead to the following relations:
\begin{eqnarray}
   & \left(L^{-4}{F_{,\,L}L_{,\,\zeta}}
    \right)_{,\,z}
    -  V^{-1}L^{-4}{F_{,\,L}L_{,\,z}V_{,\,\zeta}}
    =0\,, \label{eqkilzz}\\
   & 4 V^2\left({V^{-1}L^{-4}}{\zeta F_{,\,L}L_{,\,\zeta}}
    \right)_{,\,\zeta}
     +  \psi^2 \left(V^{-1}L^{-4}{F_{,\,L}L_{,\,z}}
    \right)_{,\,z}=0\,, \label{eqkilzu} \\
   & \left({ V^{-1} L^{-4}}{F_{,\,L}L_{,\,z}}
    \right)_{,\,\zeta}  =  0\,. \label{eqkiluua}
\end{eqnarray}
The last equation can be rewritten as
\begin{equation}\label{akiluu}
   \frac{\psi_{,\,z}}{V\psi}
    \left(\frac{F_{,\,L}}{L^3}\right)_{,\,\zeta}
  +\frac{F_{,\,L}}{L^3}
    \left(\frac{\psi_{,\,z}}{V\psi }\right)_{,\,\zeta}=  0\,.
\end{equation}
By virtue of the equations  ${^{(5)}G}^{z}_{u}=0$ and (\ref{5G^zu}), it results
in $(F_{,\,L}/{L^3})_{,\,L}=0$. This equation gives $F_{,\,L}=AL^3$ where $A $
is a constant. In further we suppose that $A=1$.

Now the equation (\ref{eqkilzz})  takes the form
\begin{equation}\label{eqkilzze}
  \left({L^{-1}}{L_{,\,\zeta}}\right)_{,\,z} -
    {L^{-1} V^{-1}}{L_{z}V_{,\,\zeta}}=0\,.
\end{equation}
This equation is satisfied by virtue of the Einstein equation
${^{(5)}G}^{z}_{u}=0$ and formula (\ref{5G^zu}). At last, the equation
(\ref{eqkilzu}) can be rewritten as follows
\begin{equation}\label{eqkilzue}
  4V^2\left(\psi^{-1} V^{-1}{\zeta\psi_{,\,\zeta}}\right)_{,\,\zeta}
    + \psi^2 \left(\psi^{-1} V^{-1}{\psi_{,\,z}}\right)_{,\,z}
    - 2V_{\zeta}=0\,.
\end{equation}
Taking into consideration (\ref{5G^55}) and (\ref{5G^uu}), one can see that
this equation is satisfied by virtue of the equations ${^{(5)}G^{z}_{z}}=0$ and
${^{(5)}G}^{u}_{u}= 0$.

Finally, substituting the expressions $F_{,\,L}=L^3 $ into Eq. (\ref{soleqkin})
and (\ref{solcomut}), we obtain the desired Killing vector of an extra symmetry
\begin{equation}\label{kilfin}
  \vec{\xi}=\psi^{-1} V^{-1}
  \left(L_{,\,u_{\epsilon}}\,\partial_{z}
  -L_{,\,z}\,u^{\mu}_{\epsilon}\,\partial_{\mu}\right)
  \qquad \left(u^{\mu}_{\epsilon} ={x^{\mu}}/{u_{\epsilon}}\right)\,.
\end{equation}

As a result, we come to the five-dimensional version of the
generalized Birkhoff theorem for spherically symmetric gravity in
four dimensions \cite{kram}, \cite{frolov}. According to the theorem,
the metric for the Lorentz-invariant five-dimensional space $V^{(5)}$
satisfying the vacuum five-dimensional Einstein equations admits an
extra Killing vector $\vec{\xi}$. The vector $\vec{\xi}$
(\ref{kilfin}) is linearly independent of the Killing vectors
$\vec{\xi}_{L} = x_{\mu}\omega^{\mu\nu}\partial_{\nu}$ of the initial
Lorentz group. It determines an extra symmetry $V^{(5)}$ that
corresponds to the Kaluza-Klein cylinder condition. Thus, the
Kaluza-Klein cylinder condition for the vacuum Lorentz-invariant
spaces of five-dimensional gravity are realized dynamically. The
proof of this theorem in the framework of two-dimensional dilaton
gravity and discussion of the related problems see in \cite{cavag}.

\section{Five-dimensional version of the mass function \\ and its
conservation law}\label{D5masfunct}

In addition to the usual conservation law for the geodesic equations
$\xi_{A}dx^{A}/^{(5)}ds =\mbox{const}$, the Killing vector
(\ref{kilfin}) $\vec{\xi}$ allows to construct a nontrivial
conservation law which will be satisfied by virtue of the
five-dimensional vacuum Einstein equations. Indeed, using the
contracted Bianchi identities $~^{(5)}G^{A}_{B\,;\,A}=0$ and Killing
equations (\ref{eqtKl}), we get the continuity equation
\begin{equation}\label{Lawg}
  P^{A}_{;\,A}=|^{(5)}g|^{-1/2}
  \left(\sqrt{|^{(5)}g|}~ P^{A}\right)_{,\,A}=\psi^{-4} V^{-1}
  \left((\psi^4 V P^{z})_{,\,z}+(\psi^4 V P^{\mu})_{,\,\mu}\right) =0
\end{equation}
for the vector
\begin{equation}\label{p^A}
  P^{A}=~^{(5)}G^{A}_{B} \xi^{B}
  =\psi^{-1}V^{-1}\left(L_{,\,u_{\epsilon}}{^{(5)}G^{A}_{z}}
  -L_{,\,z}{^{(5)}G}^{A}_{\mu}u^{\mu}_{\epsilon}\right)\,.
\end{equation}
Hence, using the components of the Einstein tensor (\ref{5G^mn}) --
(\ref{5G^zu}), we obtain
\begin{eqnarray}
 & P^{z} ={\psi^{-1}V^{-1}}\left(L_{,\,u_{\epsilon}}{^{(5)}G^{z}_{z}}
  -\epsilon L_{,\,z}{^{(5)}G^{z}_{u}}\right)\,,\label{p^zg} \\
 & P^{u} =\psi^{-1}V^{-1}\left(L_{,\,u_{\epsilon}}{^{(5)}G_{z}^{u}}
  -L_{,\,z}{^{(5)}G^{u}_{u}}\right) \label{p^mg}
\end{eqnarray}
where $P^{u}= u_{\mu}^{\epsilon}\,P^{\mu}$.

The equation (\ref{Lawg}) can be rewritten in the form
\begin{equation}\label{Lawpzu1}
    (u_{\epsilon}^3\psi^4 V P^{z})_{,\,z}
  +\epsilon(u_{\epsilon}^3\psi^4 V P^{u})_{,\,u_{\epsilon}} =0\,.
\end{equation}
with taking into account the relations (\ref{deriv}).  We will search for the
solutions of this equation as
\begin{equation}\label{mrp1}
   2u_{\epsilon}^3\psi^4 VP^{z}= 3\,G_{,\,u_{\epsilon}}\,,\qquad
   2\epsilon u_{\epsilon}^3\psi^4 VP^{u}=- 3\,G_{,\,z}
\end{equation}
where $G=G(\zeta\,,z)$ is a certain function. Hence we find
\begin{equation}\label{strp1}
  P^{z}= \frac{3G_{,\,u_{\epsilon}}}{2u_{\epsilon}^3\psi^4 V}\,,\qquad
  P^{u}= -\frac{3\epsilon G_{,\,z}}{2u_{\epsilon}^3\psi^4 V}\,.
\end{equation}
Substituting the expressions for $P^{z}$ and $P^{u}$ from (\ref{p^zg}) and
(\ref{p^mg}) into the relations (\ref{mrp1}), we obtain the equation which
links the gradient of the function $G $ with components of the Einstein tensor:
\begin{eqnarray}
   3G_{,\,u_{\epsilon}}=2L^3
    \left({^{(5)}\!G}^{z}_{z}L_{,\,u_{\epsilon}}
  -\epsilon\,{^{(5)}\!G}^{z}_{u}L_{,\,z}\right)\,,\label{g-u-z1} \\
  3G_{,\,z}= 2\,\epsilon L^3
  \left({^{(5)}G}^{u}_{u}L_{,\,z}
  -{^{(5)}G}_{z}^{u}L_{,\,u_{\epsilon}}\right)\,.\label{g-u-z2}
\end{eqnarray}
From here, with the help of Eqs.~(\ref{5G^55}) -- (\ref{5G^uu}), we obtain the
following expression $G_{,\,a} = \left(4\zeta^2\psi\psi_{\zeta}+4\zeta^3
\psi^{2}_{\zeta}-\zeta^2{\psi^2 \psi^{2}_{z}}/{V^2}\right)_{,\,a}$ whence we
find (up to an additive constant)
\begin{equation}\label{5charg}
  G =\zeta^2\left(4\psi\psi_{\zeta}+4\zeta \psi^{2}_{\zeta}
  -{\psi^2 \psi^{2}_{z}}/{V^2}\right)\,.
\end{equation}
Due to the vacuum Einstein equations (\ref{2.1}), the conservation
law $G =G(\zeta\,,z)=\mbox{const}$ follows from (\ref{g-u-z1}) --
(\ref{g-u-z2}).

Let us now rewrite the metric (\ref{pmmetr}) in the $(2+3)$ decomposed form
\begin{equation}\label{pmmetr1}
    ^{(5)}ds^2 =  {^{(2)}ds^2} -
    {L^2}\,^{(3)}d\Omega^{2}_{\epsilon}
\end{equation}
where $^{(2)}ds^2$ is the metric of space $V^{(2)}$ which is a section $y^a
=\mbox{const}$ of the space $V^{(5)}$. In the coordinates
$\{u_{\epsilon}\,,z\}$ we have
\begin{equation}\label{2metra}
   ^{(2)}ds^2 = \gamma_{ab}dx^{a}dx^{b}
   = \epsilon\psi^{2}(z,\zeta)\, du^2_{\epsilon}
    -V^{2}(z,\zeta)dz^2 \,.
\end{equation}
The metric describes subspace $\zeta=\eta_{\mu\nu}x^{\mu}x^{\nu}<0 $,
when $\epsilon=1 $ or subspace $\zeta=
\eta_{\mu\nu}x^{\mu}x^{\nu}>0$, when $\epsilon =-1$. In the
coordinates $\{\zeta={\epsilon}u_{_{\epsilon}}^2\,,z\}$ we have
\begin{equation}\label{2metrb}
   ^{(2)}ds^2 ={4^{-1}\zeta^{-1}}{\psi^{2}(z,\zeta)}\, d\zeta^2
    -V^{2}(z,\zeta)dz^2\,.
\end{equation}
This form is applicable in both subspaces $\zeta >0$ and $\zeta<0$. Let us
write out also two-dimensional representation of the Killing vector
(\ref{kilfin}) in the coordinates $\{\zeta={\epsilon}u_{_{\epsilon}}^2\,,z\}$
\begin{equation}\label{kilfin1}
  \vec{\xi}={u_{\epsilon}}{V^{-1} L^{-1}}
  \left(L_{,\,u_{\epsilon}}\,\partial_{z}
  -L_{,\,z}\,\partial_{u_{\epsilon}}\right)
  ={2\zeta }{V^{-1} L^{-1}}
  \left(L_{,\,\zeta}\,\partial_{z}
  -L_{,\,z}\,\partial_{\zeta}\right)  \,.
\end{equation}
Now the integral of motion (\ref{5charg}) can be written in the two-dimensional
invariant form
\begin{equation}\label{5mass}
  G=L^2(\nabla L)^2 -\epsilon L^2
\end{equation}
where $(\nabla L)^2 = \gamma^{ab}\nabla_{a} L \nabla_{b}L \,$,
$\nabla_{a}$ $\equiv_{;\,a}$ is a covariant derivative with respect
to the metric $\gamma_{ab}$. The function $G $ is analogical to the
known mass function of GR \cite{misner}, \cite{hernandes},
\cite{matrtinez}.

\section{Lorentz-invariant five-dimensional cosmological\\ model and
reduction of gravitational action for $V^{(5)}$}\label{5Dmodel}

From the five-dimensional generalized Birkhoff theorem it follows
that for Lorentz-inva\-riant five-dimensional vacuum spaces there is
such a coordinate frame $\{z\,,u\}$ conserving the diagonal form of
the metric (\ref{pmmetr}) in which all metric functions do not depend
on the fifth coordinate $z $ (see \ref{appen4}). Then, for the metric
(\ref{2metra}) the conservation law (\ref{5mass}) has the form
\begin{equation}\label{G0l}
  \epsilon G = u^2_{\epsilon}\left(\frac{d L}
  {d u_{\epsilon}}\right)^{2}- L^2\,.
\end{equation}
By setting $G>0$, from here we obtain $ L ={\sqrt{G}}(a u_{\epsilon}
-{\epsilon}/{a u_{\epsilon}})/{2}$ where $a>0$ is constant. Using the scale
transformation $u_{\epsilon}\rightarrow \acute{u_{\epsilon}}=au_{\epsilon}
\sqrt{G}/2$, we obtain
\begin{equation}\label{Lu1}
  \psi ={L}/{u_{\epsilon}}= 1-{G}/{4\zeta}\,.
\end{equation}
By setting $G<0$, we find the same solution.

From the equation $^{(5)}G^{u}_{u}=0$, taking into account
(\ref{5G^uu}), we have
\begin{equation}\label{VuV}
  \frac{V_{\zeta}}{V}=-\frac{2\psi_{\zeta}}{\psi}\
  \frac{\psi+\zeta\psi_{\zeta}}{\psi+2\zeta\psi_{\zeta}}
\end{equation}
whence, with the help of (\ref{Lu1}), we obtain $ V^{-1}{V_{\zeta}}
=-({G/2})({\zeta^2-G^{\,2}/16})^{-1}$. Integration of this equation leads to
the following result
\begin{equation}\label{V(u)}
  V =b\,\left|\frac{1+{G}/{4\zeta}}{1-{G}/{4\zeta}}\right|
\end{equation}
where $b>0$ is a constant which can be converted into unity via the scale
transformation of $z$. The remaining vacuum equations of five-dimensional
gravity are satisfied identically.

As a result, the metric (\ref{pmmetr}) can be written as
\begin{equation}\label{dsue}
    ~^{(5)}ds^{2} = - \left(\frac{1+{\epsilon G}/{4u_{\epsilon}^2}}
       {1-{\epsilon G}/{4u_{\epsilon}^2}}\right)^2 dz^2
 + \left(1-\frac{\epsilon G}{4u_{\epsilon}^2}\right)^2
 \left(\epsilon\, du_{\epsilon}^2
    - u_{\epsilon}^2 \,{^{(3)}\!d\Omega}_{\epsilon}^2\right)\,.
\end{equation}
The metric describes the different subdomains of the space $V^{(5)}$ for
$\epsilon= \pm 1$. Taking into account (\ref{flate-conf}), we can rewrite this
metric in the form suitable for the whole space $V^{(5)}$
\begin{equation}\label{dspm}
    ~^{(5)}ds^{2} =
       - \left(\frac{1+{ G}/{4\zeta}}
       {1-{G}/{4\zeta}}\right)^2 dz^2
 + \left(1-\frac{ G}{4\zeta}\right)^2
 \eta_{\mu\nu}dx^{\mu}dx^{\nu}\,.
\end{equation}

The physical sense of our model and the constant $G$ are clarified by the
dimensional reduction of action (\ref{act}). For this purpose we shall rewrite
the interval (\ref{dspm}) in the more general (1+4)-form
\begin{equation}\label{met}
^{(5)}ds^2={^{(5)}g}_{AB}dx^A dx^B = \acute{g}_{\mu\nu}dx^\mu dx^\nu
- V^2 dz^2\,.
\end{equation}
For spaces under consideration the variables $V,\ \acute{g}_{\mu\nu}$
do not depend on the coordinate $z $. Therefore from action
(\ref{act}) we obtain through the $(1+4)$-decomposition of scalar
curvature $^{(5)}\!R$ \cite{gladush1}, integration on $z $ and
omitting a divergence
\begin{equation}\label{act1}
     I^{(4)}=-\frac{1}{16\pi G}\int \sqrt{-\acute{g}}\
      V\,{^{(4)}\!\acute{R}}\ d^4 x
\end{equation}
where $^{(4)}\!\acute{R}$ is a scalar curvature with respect to the metric
$\acute{g}_{\mu\nu}$,  $\acute{g} = \det(\acute{g}_{\mu\nu})$, $G=\hat{G}/{L}$
is a reduced gravitational constant, $L $ is a constant with length
dimensionality.

The metric (\ref{dspm}) belongs to the following general class of metrics
\begin{equation}\label{int}
       ~^{(5)}ds^{2} =- \left(\frac{1-{\varphi}/{\sqrt{6}}}
       {1+{\varphi}/{\sqrt{6}}}\, \right)^2 dz^2 +
       {\left(1+\frac{\varphi}{\sqrt{6}}\right)^2}\
       ^{(4)}ds^{\,2}
\end{equation}
where $\varphi$ is an induced scalar field, $^{(4)}ds^{\,2}= g_{\mu\nu}dx^\mu
dx^\nu$ is the physical space-time metric of $V^{(4)}$. In this parametrization
we obtain \cite{gladush2} from action (\ref{act1})
\begin{equation}\label{4.4}
  ~^{(4)}I = -\frac{1}{16\pi G} \int d^4 x
  \sqrt{-g} \left\{\left(1-\frac{\varphi^2}
  {6}\right){^{(4)}\!R}-
  {{g}}^{\,\mu\nu}\varphi_{,\,\mu}
     \varphi_{,\nu} \right\}
\end{equation}
where $^{(4)}\!R$ is a scalar curvature with respect to the metric
$g_{\mu\nu}$, $g = \det(g_{\mu\nu})$.

The new action describes the interacting gravitational ${g}_{\mu\nu}$
and scalar $\varphi $ fields. The equations of motion of a new system
have the form
\begin{eqnarray}
 & (\Delta - \frac{1}{6}R) \varphi = 0\, ,
  \label{1.6}\\
 & G_{\mu\nu} = 4\pi t_{\mu\nu}(\varphi)\equiv
  4\pi T_{\mu\nu}(\varphi)+  \frac{1}{6}\left(G_{\mu\nu}
   -\nabla_{\mu}\nabla_{\nu}
   +{g}_{\mu\nu}\Delta\right)\varphi^{2}\, , \label{4.8}\\
 & T_{\mu\nu}(\varphi)=\frac{1}{4\pi}\left(\varphi_{\mu}\varphi_{\nu}
  -\frac{1}{2}g_{\mu\nu}(\nabla \varphi)^2\right)  \label{4.9}
\end{eqnarray}
where $t_{\mu\nu}$ is a conformally invariant energy-momentum tensor of the
scalar field $\varphi$, $\Delta =\nabla^{\mu}\nabla_{\mu}$ and all quantities
are calculated  with respect to the metric $g_{\mu\nu}$.

In our case of the Lorentz-invariant five-dimensional cosmological model with
the metric (\ref{dspm}) the scalar field and physical metric in (\ref{int})
take the form
\begin{equation}\label{phy1}
    \varphi=-\frac{\sqrt{6}\,G}{4\zeta}
    =-\frac{\sqrt{6}\,G}{4\eta_{\mu\nu}x^{\mu}x^{\nu}}\,,
    \qquad g_{\mu\nu}=\eta_{\mu\nu}\,.
\end{equation}
Hence it can be seen that the constant $G $ can be interpreted as a
charge of the scalar field $ \varphi $.

Note that the model contains the nontrivial scalar field $\varphi $
(\ref{phy1}) with vanishing conformally invariant energy-momentum tensor
$t_{\mu\nu}(\varphi)=0$. If we consider the equation
$\hat{t}_{\mu\nu}(\varphi)\equiv t_{\mu\nu} (\varphi) =\lambda\delta_{\mu\nu}
\varphi $ as the eigenvalue equations for the operator $\hat{t}_{\mu\nu}$, then
the scalar field $\varphi $ is an eigenfunction of this operator corresponding
to the zero eigenvalue. Therefore, the scalar field (\ref{phy1}) can be
interpreted as a classical zero mode of the conformally invariant
energy-momentum tensor $t_{\mu\nu}$. The flat space-time can be interpreted
also as a zero mode of the Einstein equations since it is trivial solution of
the vacuum Einstein equations. As a result, we can consider the
Lorentz-invariant solution of the Einstein equations for the interacting
gravitational and conformal scalar fields as a zero mode of GR, and the field
system (\ref{phy1}) itself as a vacuum-like configuration of GR.

If we take into account the above reasons, the indicated vacuum-like
configuration of GR can be interpreted as a manifestation of a
Lorentz-invariant mode of five-dimensional gravitation.

\section{Geometry of model, regularity conditions,\\ and compactification of the
fifth coordinate}\label{geometry}

The metric (\ref{dspm}) of the space $V^{(5)}$ has a singularity on a
light cone $\zeta\equiv\eta_{\mu\nu}x^{\mu}x^{\nu}=0$. This
singularity divides the hypersurface $z=\mbox{const}$ into two
regions: $U$-region when $\zeta>0\, (\epsilon=1)$ and $V$-region when
$\zeta <0\, (\epsilon=-1)$, where $ \epsilon =\zeta/|\zeta|$.

The singularity $\zeta=0$ has a coordinate character. It turns out that the
invariant of curvature $I= {^{(5)}\!R}_{ABCD}{^{(5)}\!R}^{ABCD}$ is finite here
(see \ref{appen3}). Therefore, the metrics (\ref{dsue}) with different
$\epsilon=\pm 1$ can be interpreted as the metrics of regions $U\subset
V^{(5)}$ and $V\subset V^{(5)}$. Then we can consider the metric (\ref{dspm})
as the analytical expansion of the metrics (\ref{dsue}) on the whole space
$V^{(5)}$.

According to (\ref{eu2}), we shall introduce coordinates in $U$- and
$V$-regions
\begin{equation}\label{uev}
  u=u_{+1}=\pm\sqrt{\eta_{\mu\nu}x^{\mu}x^{\nu}}\,,  \qquad
  v=u_{-1}=\sqrt{-\eta_{\mu\nu}x^{\mu}x^{\nu}}\,.
\end{equation}
We suppose that $-\infty<u<\infty$ and $0<v<\infty$.

Note that the region $U$ is a union of subregions of the future $U^{+} \ (u>0)$
and past $U^{-}\ (u<0)$. The subregions $U^{+}$ and $U^{-}$ are divided into
the subregions $U^{+}_{+}$, $U^{+}_{-}$ and $U^{-}_{+}$, $U^{-}_{-}$  by the
singular space-like surfaces
\begin{equation}\label{sing+}
  u^2={|G|}/{4}\,,
\end{equation}
so that $U^{+}=U^{+}_{+}\cup U^{+}_{-}$ and $U^{-}=U^{-}_{+}\cup U^{-}_{-}$,
while the region $V$ is divided into the subregions $V_{-}$ è $V_{+}$ by the
singular timelike surface
\begin{equation}\label{sing-}
  v^2={|G|}/{4}\,,
\end{equation}
so that $V=V_{-}\cup V_{+}$. Thus, we have the following subregions
of the hypersurface $z=\mbox{const}$
\begin{eqnarray}
 & U^{+}_{-} = \{u:0<u<\sqrt{|G|}/2\}\,, \quad
 & U^{+}_{+} = \{u:\sqrt{|G|}/2<u<\infty\}\,,\label{domen1}  \\
 & U^{-}_{+} = \{u:-\sqrt{|G|}/2<u<0\}\,, \
 & U^{-}_{-} = \{u:-\infty<u<-\sqrt{|G|}/2\}\,,  \label{domen2}  \\
 & V_{-} = \{v:0<v<\sqrt{|G|}/2)\}\,, \quad
 & V_{+} = \{v:\sqrt{|G|}/2<v<\infty\}\,. \label{domen3}
\end{eqnarray}
The metrics (\ref{dsue}) and (\ref{dspm}) are invariant with respect to the
inversion
\begin{equation}\label{inver+}
  u=\epsilon_{1}{G}/{4\tilde{u}}\,,\quad
 x^{\mu}=\epsilon_{1}{G\tilde{x}^{\mu}}/{\tilde{4\zeta}}\,,\quad
\tilde{\zeta}=\eta_{\mu\nu}\tilde{x}^{\mu}\tilde{x}^{\nu}>0\,,
\end{equation}
\begin{equation}\label{inver-}
v={|G|}/{4\tilde{v}}\,,\quad
x^{\mu}={|G|\tilde{x}^{\mu}}/{4\tilde{\zeta}}\,, \quad
\tilde{\zeta}=\eta_{\mu\nu}\tilde{x}^{\mu}\tilde{x}^{\nu}<0
\end{equation}
where $\epsilon_{1}=\pm 1$. These inversions induce maps of subregions
(\ref{domen1})--(\ref{domen3}):
\begin{eqnarray}
 & U^{+}_{+} \Leftrightarrow U^{+}_{-}\,, \qquad U^{-}_{-}
  \Leftrightarrow U^{-}_{+}\qquad (\epsilon_{1}G>0)\,,\label{domen-map+}\\
 & U^{+}_{+} \Leftrightarrow U_{+}^{-}\,, \qquad U^{-}_{-}
  \Leftrightarrow  U_{-}^{+}\qquad (\epsilon_{1}G>0)\,,\label{domen-map+2}\\
 & V_{+} \Leftrightarrow V_{-}\,.\label{domen-map-}
\end{eqnarray}
According to (\ref{dsue}), the $U $-region is
described in the coordinates $\{u,\,z\}$ by the metric
\begin{equation}\label{ds+u}
    ~^{(5)}ds^{2} = - \left(\frac{1+{ G}/{4u^2}}
       {1-{ G}/{4u^2}}\right)^2 dz^2
 + \left(1-\frac{ G}{4u^2}\right)^2 (du^2 - u^2 d\Omega^2_{+})\,.
\end{equation}
The point $ u=0 $ is a regular point of the metric. The singularity character
of surfaces $u^2=G/4 $ depends on a sign of $G$. In the case $ G<0$ we have a
surface where extreme curvature is reached, $I_{\max}={72}/{G^2}$, and in case
$G>0$ we have a surface of singular curvature: $\lim I \rightarrow \infty $,
when $u^2\rightarrow G/4 $ (see \ref{appen3}).

If one uses a coordinate
\begin{equation}\label{tz}
    T=u - {G}/{4u}\,,
\end{equation}
the metric (\ref{ds+u}) takes the form
\begin{equation}\label{dsT1+}
   ~^{(5)}ds^2 = -\left(1 +{G}/{T^2}\right)dz^2
    + \left(1 +{G}/{T^2}\right)^{-1}{dT^2}
   - T^2\,d\Omega^{2}_{+}\,.
\end{equation}
Hence it can be seen that coordinates $\{z,\,T\}$ in the $U$-region are
analogous to curvature coordinates in the $T$-region of the Schwarzschild
solution in GR.

The coordinate $T$ is a single-valued function of $u$, except for $u=0$ where
it has  the point of discontinuity. The inverse function $u=(T\pm\sqrt{T^2 +
G}\ )/2$ is a double-valued function of $T$ and can have a branchpoint at
$T=\sqrt{-G}$ when $G<0$. It is easy to see that in the case $G>0$ each of the
regions $U^{+}$ and $U^{-}$ in coordinates $\{z,\,u\}$ corresponds to the same
region $(-\infty<T<\infty)$ in coordinates $\{z,\,T\}$. Thus, the regions
$U^{+}$ and $U^{-}$ in coordinates $\{z,\,T\}$ are identified. At the inverse
(\ref{inver+}) the coordinate $T$ (\ref{tz}) behaves as follows
\begin{equation}\label{TinvT}
  T=-\epsilon_{1} \tilde{T}\,,\qquad
  \tilde{T}=\tilde{u}-{G}/{4\tilde{u}}\,.
\end{equation}
Therefore, in the case $\epsilon_{1}>0$ the corresponding map of regions
(\ref{domen-map+}) and (\ref{domen-map+2}) results in the change of a time
coordinate $(T)$ sign. In addition, the subregions $U^{+}_{+}$, $U^{+}_{-}$,
$U^{-}_{+}$ and $U^{-}_{-}$ exchange their places in an appropriate way. The
metric (\ref{dsT1+}) contains a curvature singularity at the instant $T=0$. The
singularity is an invariant with respect to the inversion (\ref{inver+}) and
corresponds to singular surfaces $u=\pm\sqrt{G}/2$ for the metric (\ref{ds+u}).

If $G<0$, the ranges $0<u<\sqrt{G}/2$ and $\sqrt{G}/2 <u<\infty$ of the
variable $ u$ correspond to the ranges $\infty< T<\sqrt{|G|}$ and $\sqrt{|G|}<
T<\infty$ of the variable $T$. Hence it can be seen that the regions
$U^{+}_{+}$ and $U^{+}_{-}$ are identified in coordinates $\{z,\,T\}$.
Similarly, the ranges $-\infty <u <-\sqrt{G}/2$ and $-\sqrt{G}/2 <u<0$ of the
variable $u$ correspond to the ranges $-\infty< T<-\sqrt{|G|}$ and
$-\sqrt{|G|}< T<-\infty$ of the variable $T$. Therefore, the regions
$U^{-}_{+}$ and $U^{-}_{-}$ in coordinates $\{z,\,T\}$ are identified as well.

Note that the regular surfaces $u=\pm\sqrt{|G|}/2 $ correspond to the
extreme values of time $T=\pm\sqrt{|G|}$ and the curvature invariant
$I=72/G^2$ (see \ref{appen3}). We also note that 3-spheres
$T=\pm\sqrt{|G|}\,$, $z=\mbox{const}$ (with respect to the metric
(\ref{dsT1+})) are similar to the event horizon of the Schwarzschild
metric in GR. However, a passage into the region $|T|<\sqrt{|G|}$
which is similar to a passage into the $T$-region of the
Schwarzschild solution of GR is forbidden in the classical theory.
This is because of such a change of coordinate sense that we should
perform the replacement $ z\rightarrow T \,,\ T\rightarrow z$. Then
the metric will depend on the fifth coordinate $z$, that does not
correspond to the problem set up.

Hence it follows that at a negative scalar charge the
Lorentz-invariant Universe in the KK theory exists only when
coordinate time $|T|>\sqrt{|G|}$, disappearing at $T=-\sqrt{|G|}$ and
arising at $T=\sqrt{|G|}$ with spatial section as an isotropic
expanding 3-sphere $S^3 $ with initial radius $\sqrt{|G|}$.

According to (\ref{dsue}), in coordinates $\{v,\,z\}$ the $V$-region is
described by the metric
\begin{equation}\label{ds+v}
    ~^{(5)}ds^{2} = - \left(\frac{1-{ G}/{4v^2}}
       {1+{ G}/{4v^2}}\right)^2 dz^2
 - \left(1+\frac{ G}{4v^2}\right)^2
 (dv^2 + v^2 d\Omega^2_{-})\,.
\end{equation}
The point $v=0$ is a regular point of the metric. The character of a
singularity of the surface $v=\sqrt{|G|}\,/2$ depends on a sign of $G$. In the
case $ G>0$ we have a surface where extreme curvature is reached,
$I_{\max}={72}/{G^2}$ when $v^2= -G/4$, and in the case $G<0$ we have a surface
of singular curvature $\lim I \rightarrow \infty $ when $v^2\rightarrow -G/4$
(see \ref{appen3}).

If one uses a coordinate
\begin{equation}\label{rz}
    R=v + {G}/{4v}\,,
\end{equation}
the metric (\ref{ds+v}) takes the form
\begin{equation}\label{dsT1-}
   ~^{(5)}ds^2 = -\left(1 -{G}/{R^2}\right)dz^2
    - \left(1 -{G}/{R^2}\right)^{-1}{dR^2}
   - R^2\,d\Omega^{2}_{-}\,.
\end{equation}
Hence it can be seen that coordinates $\{z,\,R\}$ in the $V$-region are an
analogous to curvature coordinates in the $R$-region of the Schwarzschild
solution in GR. The coordinate $R$ sense is a radius of the three-dimensional
pseudosphere $S(1,2)$ with the induced metric $R^2d\Omega^{2}_{-}$. This metric
is defined in the stereographic coordinates by the formula (\ref{domegapm}).

It is easy to see that in the case $G>0$ the expression in the
right-hand side of (\ref{rz}) is invariant with respect to the
inversion (\ref{inver-}), therefore the map of regions
(\ref{domen-map-}) occurs without changing the radial coordinate $R$.
Hence it follows that at the transformation (\ref{rz}) we have the
regions $V_{-}$ and $V_{+}$ overlapping, since these regions are
mapped into the same region $\sqrt{G}\leq R <\infty $. In addition,
the minimum radius $R=\sqrt{G}$ of the pseudosphere $S(1,2)$
corresponds to the regular surface $v=\sqrt{G}/2$ with the curvature
invariant extremum $I_{\max}={72}/{G^2}$ on it (see \ref{appen3}).
The pseudosphere $R =\sqrt{G} $ (with respect to the metric
(\ref{dsT1-})) is similar to the event horizon of the Schwarzschild
metric. The region $R<\sqrt{G}$ is nonphysical because of the
violation of the signature conditions. Therefore, the region
$R<\sqrt{G}$ falls out of the range of definition of the considered
space that is characteristic for a so called topological texture
\cite{gang}. In this case there are 3-spheres which are not
contractible in a point. However, in the case under consideration we
have an exotic case of the nocontractible 3-pseudospheres $S(1,2)$
instead of 3-spheres $S^3$.

If $G<0$, the expression in the right-hand side (\ref{rz}) changes a
sign at the inversion (\ref{inver-}). Therefore, the map of regions
(\ref{domen-map-}) occurs by changing a coordinate $R$ sign. Besides,
it the subregions $V_{-}$ and $V_{+}$ exchange their places. The
range $0<v<\infty$ of variable $v$ corresponds to the range $-\infty<
R <\infty$ of the variable $R$. The central singularity $R=0$ of the
metric (\ref{dsT1-}) corresponds to the surface $v=\sqrt{|G|}/2$ for
the metric (\ref{ds+v}) and is invariant with respect to the
inversion (\ref{inver-}).

Let us now consider the regularity conditions  for $V^{(5)}$ on the
``\,horizon'' $R=\sqrt{G}$ for (\ref{dsT1-}) that corresponds
$v=\sqrt{G}/2\ (G>0)$ for (\ref{ds+v}). With this end in view we
consider the induced metric
\begin{equation}\label{5/2ds+v}
    dl^{2}= \psi^{2}(v)\,dv^2 + V^{2}(v)dz^2
    =  \left(1+\frac{ G}{4v^2}\right)^2 dv^2
    + \left(\frac{1-{ G}/{4v^2}}
       {1+{ G}/{4v^2}}\right)^2 dz^2 \,
\end{equation}
on the surface $V^{(2)}\subset V^{(5)}$: $y^i=\mbox{const}$ when
$d\Omega^2_{-}=0$ in the metric (\ref{ds+v}). The space $V^{(2)}$ with the
metric ({\ref{5/2ds+v}}), as locally Euclidean, can be locally embedded into
$R^{(3)}$ as a surface of revolution (see, for example, \cite{fronsdal}). Hence
it follows that the space $V^{(2)}$ is a closed space with respect to $z$.

This surface radius $$r_{v} \sim V(v)=\frac{1-{ G}/{4v^2}}{1+{
G}/{4v^2}}\qquad (\sqrt{|G|}/2\leq v<\infty)\,$$ is a monotonically
decreasing quantity as $v$ decreases with the range $\infty >
v>\sqrt{|G|}/2$ and $r_{v}=0$ when $ v=\sqrt{G}/2\ (G>0)$. Thus, we
have a conical surface of revolution. With the help of transformation
\begin{equation}\label{zru}
  r =r(v) = \int\frac{\psi(v)d v}{V(v)} = v +\frac{G}{4v}
    + \sqrt{G}\,\ln\left(\frac{2v-\sqrt{G}}{2v+\sqrt{G}}\right)
\end{equation}
conserving a symmetry the metric (\ref{5/2ds+v}) can be reduced to a
conformally flat form $dl^{2}= V^2(r)(dr^2 + dz^2)$. The additional
transformation $r=\sqrt{G}\,\ln\rho $ leads to the following metric
\begin{eqnarray}\label{2dl2a}
    dl^{2} =W^{2}(\rho)(d\rho^2 +\rho^2d\alpha^2)
    =  Ge^{-(2v/\sqrt{G}+\sqrt{G}/2v)}\,
       \frac{\left(1+\sqrt{G}/2v\right)^4}{\left(1+G/4v^2\right)^2}\,
       (d\rho^2 +\rho^2d\alpha^2)
\end{eqnarray}
where $\alpha=z/\sqrt{G}\ (G>0)$. The introduced here function $\rho$
and conformal factor $W(\rho)$ are already positive and analytical.
In order to avoid a conical singularity at the ``horizon'' $R=\sqrt{
G}$ (or $\rho=0$), $\alpha$ must have the period $2\pi$, this
implying that the extra coordinate $z$ must be periodic with the
range of changing $0\leq z \leq 2\pi\sqrt{ G}$.

From this analysis it is apparent that for the section of the space
$V^{(5)}$ at $y^i =\mbox{const}\ (i=0,1,2)$ the $v$-$z$ surface has
the shape of a semi-infinite conic surface of revolution when $G>0$.
This conical surface is smoothly rounded in its vertex and is open at
infinity; thus, it is topologically equivalent to $R^2$. Therefore,
the $V$-region has a structure $R^2 \times S(1,2)$ with an isotropic
throat of the radius $R=\sqrt{G} \ (G>0)$ as an interior boundary.
The above conic surface radius is $r_{R}= \sqrt{G(1 -{G}/{R^2})}$
with the range $0<r_{R}<r_{\infty} =\sqrt{G}$ when
$\sqrt{G}<R<\infty$. The maximum $r_{R}$ can be interpreted as a
compactification radius $r_{c}=r_{\infty}=\sqrt{G}$ of the space
$V^{(5)}$. We see that what can be called a spontaneous
compactification takes place here. Note that another mechanism of the
spontaneous compactification of $V^{(5)}$ was considered in
\cite{david} where $z$-periodicity was an effect induced by a moving
tachyon.

Since $U$- and $V$- regions are subregions of $V^{(5)}$ with a common
smooth boundary $\zeta\equiv\eta_{\mu\nu}x^{\mu}x^{\nu}=0$, the
condition $z=\alpha\sqrt{G}\ (0\leq \alpha < 2\pi)$ should be
satisfied for the whole space $V^{(5)}$. Therefore, when $G>0$ and in
accordance with (\ref{dspm}), we can write
\begin{equation}\label{ds1}
    ~^{(5)}ds^{2} =
       - \left(\frac{1+{ G}/{4\zeta}}
       {1-{ G}/{4\zeta}}\right)^2 G d\alpha^2
 + \left(1-\frac{ G}{4\zeta}\right)^2
 \eta_{\mu\nu}dy^{\mu}dy^{\nu}\,.
\end{equation}
This metric is a Lorentz-invariant mode of the equations of
five-dimensional gravity for which both cylindricity and
compactification of the fifth dimension are fulfilled dynamical\-ly.
Therefore the space under consideration is the KK space.

As already has been noted, after the dimensional reduction the model
is reduced to the vacuum-like configuration of GR for the interacting
gravitational and conformal scalar fields. Hence this vacuum-like
configuration of GR can be considered as a manifestation of a
Lorentz-invariant mode of five-dimensional gravity for which the
Kaluza-Klein postulates are realized dynamically.

\section{Reduction of action for a geodesic line
 on $V^{(5)}$}\label{geodesic}

The geodesic equations in $V^{(5)}$ follow from the action principle
$\delta^{(5)}S=0$ for an action
\begin{equation}\label{action}
  ~^{(5)}S=-\int{^{(5)}ds}=\int {^{(5)}L}d\lambda
\end{equation}
where $\lambda $ is a parameter on the world line. A Lagrangian
${~^{(5)}L}$ for the metric (\ref{metr}) has the form
\begin{equation}\label{lagr}
{~^{(5)}L} = -
    \sqrt{\psi^{2}\eta_{\mu\nu}\dot{x}^{\mu}\dot{x}^{\nu}-V^{2}\dot{z}^2}
\end{equation}
where $\dot{x}^{\mu}=d{x}^{\mu}/d\lambda$.

Let us note that the coordinate $z$ is cyclical and can be eliminated from the
Lagrangian ${^{(5)}\!L}$. For this purpose we introduce the Routh function
$\cal{R}$ \cite{haar}:
\begin{equation}\label{raus}
  {\cal R}=g\dot{z}-{^{(5)}\!L}=
  - {\psi^{2}\eta_{\mu\nu}\dot{x}^{\mu}\dot{x}^{\nu}}/{L}
\end{equation}
where
\begin{equation}\label{graus}
  g={\partial\, {^{(5)}\!L}}/{\partial\dot{z}}
  =-{V^2\dot{z}}/{L}=\mbox{const}\,
\end{equation}
is an integral of motion. The Routh function $\cal{R}$ is a Lagrange
function with respect to the coordinates ${x}^{\mu}$ and is a
Hamiltonian function with respect to the coordinate $z$. Therefore,
if only trajectories in the space-time $V^4$ are of interest for us,
it is possible to consider the function $\cal{R}$ as a reduced
Lagrangian $~^{(4)}L$ containing $g$ as s parameter.

Excluding the velocity $\dot{z}$ from the Routh function (\ref{raus}) with the
help of (\ref{graus}), we obtain
\begin{equation}\label{rausz}
  {~^{(4)}\!L}={\cal R}=\psi\sqrt{\left(1+{g^2}{V^{-2}}\right)
  \eta_{\mu\nu}\dot{x}^{\mu}\dot{x}^{\nu}}\,.
\end{equation}
Taking into account (\ref{ds1}), we find the Lagrangian of a test
particle
\begin{equation}\label{lagraus}
  {~^{(4)}\!L}=H(\zeta)\sqrt{ \eta_{\mu\nu}\dot{x}^{\mu}\dot{x}^{\nu}}
\end{equation}
where
\begin{equation}\label{hinds}
  H(\zeta)=\left(1-\frac{ G}{4\zeta}\right)
  \sqrt{1+g^2\left(\frac{1-{ G}/{4\zeta}}
  {1+{ G}/{4\zeta}}\right)^2}\,.
\end{equation}
Using the Routh function (\ref{raus}), we can rewrite the five-dimensional
action (\ref{raus}) as
\begin{equation}\label{action1}
  ~^{(5)}S=\int\! {^{(5)}L}d\lambda=\int(g d z-{\cal R}d\lambda)\,.
\end{equation}
Hence, omitting a total differential and taking into account
(\ref{rausz}) and (\ref{lagraus}), we come to the four-dimensional
Lorentz-invariant action
\begin{equation}\label{4action}
  ~^{(5)}S\rightarrow~^{(4)}S=-\int{\cal R}d\lambda=-\int {^{(4)}\!L}
  d\lambda=-\int H(\zeta) ds_{0}\,.
\end{equation}
The four-dimensional equations of motion
\begin{equation}\label{eqtmotio}
   \frac{d}{ds_{0}}\left(H\frac{dx^\mu}{ds_{0}}\right)
    =\eta^{\mu\nu}\frac{\partial H}{\partial x^\nu}\,
\end{equation}
follow from the constructed action where
$ds^2_{0}=\eta_{\mu\nu}dx^{\mu}dx^{\nu}$.

The obtained action can be interpreted as an action for a particle of
a variable mass $m= H(\zeta)$ in the flat space-time. In the case
$|\zeta|\gg0$ it is reduced to an action for a free test particle of
mass $m =\sqrt {1+g^2} $ in special relativity. In the case
$\zeta\sim G/4$ we have accelerated motion and in the case
$\zeta\rightarrow G/4 $ we deal with ultrarelativistic motion. Let us
suppose that the compactification radius of $V^{(5)}$ has an order of
Planck length $L_{Pl}$: $ r_{c}=\sqrt{G}\sim L_{Pl} $. Then the
effects of five-dimensional gravity of the considered type can show
themselves only at the Planck scales.

\section{Conclusion}\label{conclus}

The Lorentz-invariant configuration of a  scalar field with conformal
connection const\-ructed in the paper has the vanishing conformally
invariant energy-momentum tensor on the flat space-time background.
Therefore, the obtained solution is interpreted as a vacuum-like
configuration in GR without a cosmological constant.

On the other hand, it is shown that the Lorentz-invariant solution of
equations of the KK theory can be considered as a Lorentz-invariant
mode of equations of the five-dimensional gravity (without external
sources) in which the cylindricity condition and closedness with
respect to the extra coordinate $z$ are realized dynamically. After
the dimensional reduction and appropriate conformal mapping the model
is reduced to the above vacuum configuration of GR with the
conformally invariant scalar field.

Therefore, the vacuum in GR can be interpreted as a manifestation of
the Lorentz-invariant empty space of the KK theory which in turn is
the Lorentz-invariant mode in the five-dimensional gravity. The
approach discussed above, as well as the standard theory, leads to
the one-parameter family of vacuum configurations. The parameter is a
charge of the scalar field $G $ associated with the geometrical
parameter of the model, that is the compactification radius of
$r_{c}=\sqrt{G}$ of the Kaluza-Klein space $V^{(5)}$. In addition a
five-dimensional action for a geodesic line in $V^{(5)}$ is reduced
to a four-dimensional action by the dimensional reduction. We can
interpret the obtained four-dimensional effective action as an action
for a particle of the variable mass $m=H(\zeta)$ in the flat
space-time.

The offered approach is based on the five-dimensional version of the
generalized Birkhoff theorem proved in the paper. According to the
theorem, the Lorentz-invariant metric $ ^{(5)}\!g_{AB}$ satisfying
the vacuum five-dimensional Einstein equations admits the extra
Killing vector $\vec{\xi}$. This vector defines an extra symmetry of
the five-dimensional Lorentz-invariant vacuum space $V^{(5)} $
corresponding to the Kaluza-Klein cylinder condition. The
conservation law $ G=L^2(\nabla L)^2 -\epsilon L^2 = \mbox{const}$
corresponding to this symmetry is similar to the mass function for
the spherically-symmetric gravitational fields in GR. After the
dimensional reduction of the five-dimensional Hilbert action and
corresponding conformal mapping, $G $ is identified with a charge of
the conformally invariant scalar field. Therefore, the function $G$
can be called a charge function, and the conservation law corresponds
to the charge conservation for the conformally invariant scalar
field.

The regularity condition of the space $V^{(5)}$ for the obtained
metric implying the conical singularity absence leads to the  closure
condition of $V^{(5)}$ with respect to the fifth coordinate $z$ with
the period $L=2\pi\sqrt{G} $. The metric of the surface
$V^{(2)}\subset V^{(5)}$ $(y^i =\mbox{const}, i=0,1,2)$ describes a
semi-infinite conic surface of revolution with the smoothly rounded
vertex. This conic surface radius is $r_{R}= \sqrt{G(1 -{G}/{R^2})}$
with the range $0<r_{R}<r_{\infty} =\sqrt{G}$ while
$\sqrt{G}<R<\infty$. The maximum magnitude of $r_{R}$ which is
reached at $R =\infty $ can be interpreted as a compactification
radius $r_{c}=r_{\infty}= \sqrt{G}$ of the five-dimensional space
$V^{(5)}$. Finally we note that the considered five-dimensional space
for sufficiently large $\zeta=\eta_{\mu\nu}x^{\mu}x^{\nu} $ becomes a
flat space and its manifestation, the physical space-time, becomes a
usual empty space-time of special relativity. Thus, the constructed
five-dimensional cosmological model describes the mechanism of a
dynamic spontaneous compactification of the five-dimensional space
and its effective manifestation as a vacuum-like configuration of GR
containing the scalar field with a vanishing conformally invariant
energy-momentum tensor on the Minkowski space-time background.

In a sense, the Lorentz-invariant Kaluza-Klein spaces can be
understood as the limiting case of spaces of the five-dimensional GR
admitting maximally symmetric Lorentz-inva\-riant space-time
sections. We consider the obtained solution as the ``lowest''
nontrivial state of the five-dimensional geometry, which manifest
themselves in the form of the vacuum-like configuration GR. Its
nonsymmetric perturbations generated by classical or quantum
excitations can be treated as induced matter \cite{wes}. But we must
first study the problem of the stability of the space $V^{(5)}$ under
small perturbations of the type $\delta g_{AB}e^{nz}$ by analogy with
the Schwarzschild metric. Then these perturbations for stable modes
can be interpreted as matter induced by the five-dimensional
geometry.

\begin{acknowledgments}
I would like to thank E. Minguzzi for drawing my attention to the
paper \cite{karl}. Additional thanks are addressed to D. Vassilevich
for the indication on papers \cite{cavag}. At last, I am grateful to
S. Lyagushyn for his assistance in preparing the English version of
the paper.
\end{acknowledgments}

\appendix
\section{Generalized stereographic coordinates}\label{appen1}
The generalized stereographic coordinates for three-dimensional
``angular'' parts of the Minkowski metric can be introduced via using
a generalized stereographic projection \cite{gladush7},  \cite{karl}.
Taking into account definition $u^2_{\epsilon}=
\epsilon\eta_{\mu\nu}x^{\mu}x^{\nu}$, we have
\begin{equation}\label{uem}
  u_{\mu}^{\epsilon}=\epsilon\frac{\partial u_{\epsilon}}{\partial x^{\mu}}
  =\frac{\eta_{\mu\nu}x^{\nu}}{u_{\epsilon}}
  =\frac{x_{\mu}}{u_{\epsilon}}\,.
\end{equation}
This 4-vector is time-like ($\eta^{\mu\nu}u^{\epsilon}_{\mu}
u^{\epsilon}_{\nu}=1$) when $\epsilon=1$ and space-like
($\eta^{\mu\nu}u^{\epsilon}_{\mu} u^{\epsilon}_{\nu}=-1$) when $\epsilon=-1$.
The second derivative of $u_{\epsilon}$:
\begin{equation}\label{uemn}
  \epsilon\frac{\partial^2 u_{\epsilon}}{\partial x^{\mu}\partial x^{\mu}}
  =\frac{\partial u^{\epsilon}_{\mu}}{\partial x^{\mu}}
  =-\frac{1}{u_{\epsilon}}\,h_{\mu\nu}\,
\end{equation}
leads to the induced metric
\begin{equation}\label{uhmn}
  h_{\mu\nu}=\epsilon u^{\epsilon}_{\mu}u^{\epsilon}_{\nu}
  -\eta_{\mu\nu}\,, \qquad u_{\epsilon}^{\mu}h_{\mu\nu}=0
\end{equation}
defined on the surface $u^2_{\epsilon}=
\epsilon\eta_{\mu\nu}x^{\mu}x^{\nu} =\mbox{const}$. Hence it follows
  ${\partial u_{\epsilon}^{\mu}}/{\partial x^{\mu}}
  = u_{\epsilon ,\,\mu}^{\mu}
  ={3}/{u_{\epsilon}}$.

We introduce ``degenerated'' angular coordinates $u_{\epsilon}^{\mu}$
\begin{equation}\label{nm}
   u_{\epsilon}^{\mu}={x^{\mu}}/{u_{\epsilon}}
   = \eta^{\mu\nu}u^{\epsilon}_{\mu}\,.
\end{equation}
The coordinates $\{u_{\epsilon}^{\mu}\}=\{u_{+}^{\mu}\,,
u_{-}^{\mu}\}$ satisfy to relations
\begin{equation}\label{ume03}
  \eta_{\mu\nu}u_{\epsilon}^{\mu}u_{\epsilon}^{\nu}=(u^{0}_{\epsilon})^2
 -(u^{1}_{\epsilon})^2 -\delta_{ab}u^a_{\epsilon}u^b_{\epsilon}=\epsilon
 \quad (a,b = 2,3) \,,
\end{equation}
and $\eta_{\mu\nu}u^{\mu}_{\epsilon}dx^{\nu}=du_{\epsilon}\,, \
\eta_{\mu\nu}u^{\mu}_{\epsilon}du^{\nu}_{\epsilon}=0 $. Using these formulae
and (\ref{uhmn}), we obtain for the Minkowski metric
\begin{equation}\label{ds0+}
  ^{(4)}ds^{2}_{0} =\epsilon du^2_{\epsilon} -
  h_{\mu\nu}dx^{\mu}dx^{\nu} =\epsilon du^2_{\epsilon}
  - u^2_{\epsilon} \,{^{(3)}\!d\Omega}^2_{\epsilon}
\end{equation}
where
\begin{equation}\label{domega+z}
  \,{^{(3)}\!d\Omega}^2_{\epsilon} =
  h_{\mu\nu}du^{\mu}_{\epsilon}du^{\nu}_{\epsilon}=-
  (\eta_{\mu\nu}du^{\mu}_{\epsilon}du^{\nu}_{\epsilon})_{|u^2_{\epsilon}
  ={const}}
\end{equation}
are induced metrics on the surfaces $\eta_{\mu\nu}u_{\epsilon}^{\mu}
u_{\epsilon}^{\nu}=\epsilon$. They represents ``angular parts'' of
the Minkowski metrics inside and outside the light cone.

Now we proceed from the degenerated coordinates $u^{\mu}_{\epsilon}$ to the
nondegenerate generalized stereographic coordinates $y^i$ by means of the
formulae
\begin{equation}\label{anglee}
    u^a_{\epsilon} =
    \frac{y^a}{1-\epsilon S_{\epsilon}^2/4}\,,\qquad
  u^1_{+} = \frac{y^1}{1-S_{+}^2/4}\,, \qquad
  u^0_{-} = \frac{y^1}{1+S_{-}^2/4}
\end{equation}
where
 $ S_{\epsilon}^2=\epsilon(y^1)^2 +(y^2)^2 +(y^3)^2 $.
Then, according to (\ref{ume03}), we have
\begin{equation}\label{u0+}
  u^0_{+} = \frac{1+S_{+}^2/4}{1-S_{+}^2/4}\,,\qquad
  u^1_{-} = \frac{1-S_{-}^2/4}{1+S_{-}^2/4}\,.
\end{equation}
Now the angular part of the Minkowski metrics (\ref{domega+z}) takes
the form (\ref{domegapm}). This metric describes the
three-dimensional space of constant curvature $(K_{0}=-\epsilon)$.

\section{The curvature tensor, the Ricci tensor and\\ curvature scalar of
$V^{(5)}$ }\label{appen3}

In order to find the curvature tensor of space $V^{(5)}$ with the
metric (\ref{metr}) at first we shall write out derivatives of the
variable $\zeta=\eta_{\mu\nu}x^{\mu}x^{\nu}$:
\begin{equation}\label{zetam}
\zeta_{,\,\mu}=
  \frac{\partial \zeta}{\partial x^{\mu}}
    =2\eta_{\mu\nu}x^{\nu}=2x_{\mu}\,, \quad
    \zeta_{,\,\mu\nu}=
  \frac{\partial^2 \zeta}{\partial x^{\mu}\partial x^{\mu}}
  =2\eta_{\mu\nu}\,.
\end{equation}
Taking into account these formulae, we obtain the Christoffel symbols
for the metric (\ref{metr})
\begin{equation}\label{g555-m55}
  \Gamma^{5}_{55}=\frac{V_{z}}{V}\,,\quad
  \Gamma^{5}_{5\mu}=\frac{2V_{\zeta}}{V}\,x^{\mu}\,,\quad
   \Gamma^{\mu}_{55}=\frac{2VV_{\zeta}}{\psi^2}\,x^{\mu}\,,\quad
  \Gamma^{5}_{\mu\nu}=\frac{\psi\psi_{z}}{V^2}\,\eta_{\mu\nu}\,,
\end{equation}
\begin{equation}\label{gmnr}
  \Gamma^{\mu}_{\nu 5}=\frac{\psi_{z}}{\psi}\,\delta^{\mu}_{\nu}\,,\qquad
  \Gamma^{\mu}_{\nu\rho}=\frac{2\psi_{\zeta}}{\psi}\,
  (\delta^{\mu}_{\nu}x_{\rho}+\delta^{\mu}_{\rho}x_{\nu}-\eta_{\nu\rho}x^{\mu})
\end{equation}
where the indexes ``$_{\zeta}$`` and ``$_{z}$'' denote the partial
derivatives $ \partial/\partial {\zeta} $ and $\partial/\partial{z}$,
accordingly.

The components of the curvature tensor for the metric (\ref{metr})
have the form
\begin{eqnarray}
  ^{(5)}R^{z}_{\mu\, z\,\nu}
 & = & \left(\frac{\psi\psi_{zz}}{V^2}
   -\frac{\psi\psi_{z}V_{z}}{V^3}
  -\frac{4\zeta\psi_{\zeta}V_{\zeta}}{\psi V}
  -\frac{2V_{\zeta}}{V}\right)\eta_{\mu \nu}
  +  4\left( -\frac{V_{\zeta\zeta}}{V}
    +\frac{2\psi_{\zeta}V_{\zeta}}{\psi V}\right)x_{\mu}x_{\nu}\,,\label{R^5-m5m}\\
  ^{(5)}R^{\mu}_{\nu\,\rho\, z} & = & 2\left(\frac{\psi_{z\zeta}}{\psi}
  -\frac{\psi_{z}\psi_{\zeta}}{\psi^2}
  - \frac{\psi_{z}V_{\zeta}}{\psi V}\right)
  \left(x^{\mu}\eta_{\nu\rho} -
  x_{\nu}\delta^{\mu}_{\rho}\right)\,,\label{R^m-nr5}\\
    ^{(5)}R^{\mu}_{\nu\,\rho\,\sigma}
 & = & \left(\frac{4\zeta\psi^{2}_{\zeta}}{\psi^2}
  +\frac{4\psi_{\zeta}}{\psi} -\frac{\psi^2_{z}}{V^2}\right)
  \left(\delta^{\mu}_{\sigma}\eta_{\nu\rho}
   -\delta^{\mu}_{\rho}\eta_{\nu\sigma}\right) + \nonumber \\
 & + & 4\left(\frac{\psi_{\zeta\zeta}}{\psi}
    -\frac{2\psi^{2}_{\zeta}}{\psi^2}\right)
    \left(x^{\mu}(x_{\sigma}\eta_{\nu\rho}
   -x_{\rho}\eta_{\nu\sigma})- x_{\nu}(x_{\sigma}\delta^{\mu}_{\rho}
   -x_{\rho}\delta^{\mu}_{\sigma}) \right)\,.\label{R^m-nrs}
\end{eqnarray}
Hence we find the components of the Ricci tensor and curvature scalar
\begin{eqnarray}
&& ^{(5)}R_{zz}= -\frac{4V^2}{\psi^2}\left(
  \frac{\psi\psi_{zz}}{V^2}
   -\frac{\psi\psi_{z}V_{z}}{V^3}
  -\frac{2\zeta\psi_{\zeta}V_{\zeta}}{\psi V} -\frac{2V_{\zeta}}{V}
  -\frac{\zeta V_{\zeta\zeta}}{V} \right)\,,\label{R55}\\
 && ^{(5)}R_{\mu z} = -6\left(\frac{\psi_{z\zeta}}{\psi}
  -\frac{\psi_{z}\psi_{\zeta}}{\psi^2}
  - \frac{\psi_{z}V_{\zeta}}{\psi V}\right) x_{\mu}\,,\label{Rm5}\\
   && \qquad ^{(5)}R_{\mu\nu}=
    4\left(-\frac{2\psi_{\zeta\zeta}}{\psi}
    +\frac{4\psi^{2}_{\zeta}}{\psi^2}
     -\frac{V_{\zeta\zeta}}{V}
    +\frac{2\psi_{\zeta}V_{\zeta}}{\psi V}\right)x_{\mu}x_{\nu}+ \\ \label{Rmn}
   && + \left(\frac{\psi\psi_{zz}}{V^2}
   -\frac{\psi\psi_{z}V_{z}}{V^3} +\frac{3\psi^2_{z}}{V^2}
  - \frac{4\zeta\psi_{\zeta\zeta}}{\psi}
   -\frac{4\zeta\psi^{2}_{\zeta}}{\psi^2}-\frac{12\psi_{\zeta}}{\psi}
    -\frac{4\zeta\psi_{\zeta}V_{\zeta}}{\psi V}
  -\frac{2V_{\zeta}}{V}\right)\eta_{\mu\nu} \,,  \nonumber
\end{eqnarray}
\begin{eqnarray}\label{5R}
 ^{(5)}R=\frac{4}{\psi^2}\left(
  \frac{2\psi\psi_{zz}}{V^2}+\frac{3\psi^2_{z}}{V^2}
   -\frac{2\psi\psi_{z}V_{z}}{V^3}
  -\frac{2\zeta V_{\zeta\zeta}}{V}-\frac{4V_{\zeta}}{V}
  -\frac{6\zeta\psi_{\zeta\zeta}}{\psi}
  -\frac{12\psi_{\zeta}}{\psi}
  -\frac{4\zeta\psi_{\zeta}V_{\zeta}}{\psi V}\right) \,. 
\end{eqnarray}
At last we find components of the Einstein tensor
\begin{eqnarray}
  ^{(5)}G_{zz} & = & -\frac{6V^2}{\psi^2}\left(
  \frac{2\zeta\psi_{\zeta\zeta}}{\psi}
  +\frac{4\psi_{\zeta}}{\psi}-\frac{\psi^2_{z}}{V^2}
  \right)\,,\label{5G55}\\
  ^{(5)}G_{\mu z} & = &  ^{(5)}R_{\mu z}
  = -6\left(\frac{\psi_{z\zeta}}{\psi}
  -\frac{\psi_{z}\psi_{\zeta}}{\psi^2}
  - \frac{\psi_{z}V_{\zeta}}{\psi V}\right) x_{\mu}\,,\label{5G5m}\\
     ^{(5)}G_{\mu\nu} & = &
     4\left(-\frac{2\psi_{\zeta\zeta}}{\psi}
    +\frac{4\psi^{2}_{\zeta}}{\psi^2}
     -\frac{V_{\zeta\zeta}}{V}
+\frac{2\psi_{\zeta}V_{\zeta}}{\psi V}\right)x_{\mu}x_{\nu}-\nonumber
\\
    -  \left(
  \frac{3\psi\psi_{zz}}{V^2}+\frac{3\psi^2_{z}}{V^2}\right.
   & - & \left.\frac{3\psi\psi_{z}V_{z}}{V^3}
  -\frac{4\zeta V_{\zeta\zeta}}{V}
   - \frac{6V_{\zeta}}{V}  
  -  \frac{8\zeta\psi_{\zeta\zeta}}{\psi}
  +\frac{4\zeta\psi^{2}_{\zeta}}{\psi^2}
  -\frac{12\psi_{\zeta}}{\psi}
-\frac{4\zeta\psi_{\zeta}V_{\zeta}}{\psi
V}\right)\eta_{\mu\nu}\,.\label{5Gmn}
\end{eqnarray}
Further we introduce the induced metric (\ref{uhmn}) and projection
operator
 $ P^{\mu}_{\nu}=- h^{\mu}_{\nu}= -\epsilon u_{\epsilon}^{\mu}u^{\epsilon}_{\nu}
  +\delta^{\mu}_{\nu}$
where $ u_{\epsilon}^{\mu}P_{\mu}^{\nu}=0$, $
P_{\mu}^{\nu}P_{\nu}^{\rho}=P_{\mu}^{\rho}$. Here
$\epsilon={\zeta}/{|\,\zeta|}$,
$\zeta=\eta_{\mu\nu}x^{\mu}x^{\nu}=\epsilon u^2_{\epsilon}$,
$u^{\mu}_{\epsilon}=\eta^{\mu\nu}u_{\nu}^{\epsilon}$, and $
u_{\mu}^{\epsilon}u^{\mu}_{\epsilon}=\epsilon$. Using these formulae,
we can rewrite components of the curvature tensor (\ref{R^5-m5m}),
(\ref{R^m-nr5}), and (\ref{R^m-nrs}) in (1+1+3)-decomposed form
\cite{gladush1}
\begin{eqnarray}\label{hR^5-m5m}
 ^{(5)}R^{z}_{\mu\, z\,\nu} & = & \epsilon
 \left( \frac{\psi\psi_{zz}}{V^2}
   -\frac{\psi\psi_{z}V_{z}}{V^3}
    -\frac{4\zeta V_{\zeta\zeta}}{V}
    +\frac{4\zeta\psi_{\zeta}V_{\zeta}}{\psi V}
   -\frac{2V_{\zeta}}{V}\right)u_{\mu}^{\epsilon}u_{\nu}^{\epsilon}
   + \nonumber \\
 & +& \left(-\frac{\psi\psi_{zz}}{V^2}
   +\frac{\psi\psi_{z}V_{z}}{V^3}
  +\frac{4\zeta\psi_{\zeta}V_{\zeta}}{\psi V}
  +\frac{2V_{\zeta}}{V}\right)h_{\mu\nu}\,,
\end{eqnarray}
\begin{equation}\label{hR^m-nr5}
  ^{(5)}R^{\mu}_{\nu\,\rho\, z} =
  2\psi^{-1}\sqrt{\epsilon\zeta}\,\left({\psi_{z\zeta}}
  -{\psi^{-1}\psi_{z}\psi_{\zeta}}
  - {\psi_{z}V^{-1}V_{\zeta}}\right)
  \left(u_{\nu}^{\epsilon} h^{\mu}_{\rho}
  -u^{\mu}_{\epsilon}h_{\nu\rho}
  \right)\,,
\end{equation}
\begin{eqnarray}\label{hR^m-nrs}
 && ~ \qquad \qquad ^{(5)}R^{\mu}_{\nu\,\rho\,\sigma}
  =\left(\frac{4\zeta\psi^{2}_{\zeta}}{\psi^2}
  +\frac{4\psi_{\zeta}}{\psi}-\frac{\psi^2_{z}}{V^2} \right)
  \left(h^{\mu}_{\sigma}h_{\nu\rho}
   -h^{\mu}_{\rho}h_{\nu\sigma}\right) +  \\
 && +\, 4\epsilon\left(\frac{\zeta\psi_{\zeta\zeta}}{\psi}
    -\frac{\zeta\psi^{2}_{\zeta}}{\psi^2}+\frac{\psi_{\zeta}}{\psi}
    -\frac{\psi^2_{z}}{4V^2}\right)
    \left(u^{\mu}_{\epsilon}(u^{\epsilon}_{\rho}h_{\nu\sigma}
   -u^{\epsilon}_{\sigma}h_{\nu\rho})-
   u^{\epsilon}_{\nu}(u^{\epsilon}_{\rho}h^{\mu}_{\sigma}
   -u^{\epsilon}_{\sigma}h^{\mu}_{\rho}) \right)\,. \nonumber
\end{eqnarray}
Then components of the Einstein tensor (\ref{5G55}), (\ref{5G5m}),
and (\ref{5Gmn}) take the form (\ref{5G^mn})- (\ref{5G^hh}). In these
relations we use the formulae $ A_{\mu }= A_{u}u_{\mu}^{\epsilon}
\,,\, A^{\mu }=\epsilon A^{u }u^{\mu}_{\epsilon}$, so that $A_{\mu}
A^{\mu}=A_{u }A^{u }=\epsilon\psi^2(A^{u})^2$. In the last expression
the index operations are performed by rules
\begin{equation}\label{u^-uac}
  A_{u}=\epsilon\psi^2 A^{u}\,,\quad
  A^{u}=\epsilon\psi^{-2} A_{u}\,.
\end{equation}

Further, by virtue of the conservation law (\ref{5charg}), from
(\ref{hR^m-nrs}) we have
\begin{eqnarray}\label{5hR^m-nrs}
 && \qquad \qquad \qquad   ^{(5)}R^{\mu}_{\nu\,\rho\,\sigma}
    =\frac{G}{\zeta^2\psi^2}
     \left(h^{\mu}_{\sigma}h_{\nu\rho}
   -h^{\mu}_{\rho}h_{\nu\sigma}\right) +  \\
 && +\, 4\epsilon\left(\frac{\zeta\psi_{\zeta\zeta}}{\psi}
    -\frac{\zeta\psi^{2}_{\zeta}}{\psi^2}+\frac{\psi_{\zeta}}{\psi}
    -\frac{\psi^2_{z}}{4V^2}\right)
    \left(u^{\mu}_{\epsilon}(u^{\epsilon}_{\rho}h_{\nu\sigma}
   -u^{\epsilon}_{\sigma}h_{\nu\rho})-
   u^{\epsilon}_{\nu}(u^{\epsilon}_{\rho}h^{\mu}_{\sigma}
   -u^{\epsilon}_{\sigma}h^{\mu}_{\rho}) \right)\,.\nonumber
\end{eqnarray}
In our case the five-dimensional Einstein equations (\ref{2.1}), with
taking into account (\ref{5G^55})--(\ref{5G^hh}), can be written as
\begin{equation}\label{5pszz}
    \zeta\psi_{\zeta\zeta}= - 2\psi_{\zeta}\,,
\end{equation}
\begin{equation}\label{5Vza3}
  \frac{V_{\zeta}}{V} =
 -\frac{2\psi_{\zeta}}{\psi}\left(\frac{\zeta\psi_{\zeta}+\psi}
 {2\zeta\psi_{\zeta}+\psi}\right) \,,
\end{equation}
\begin{equation}\label{5Vzz}
    \zeta V_{\zeta\zeta} =
    -{2V_{\zeta}}{\psi^{-1}}\left(\zeta\psi_{\zeta}+\psi\right)\,.
\end{equation}
Further, for components of the curvature tensor (\ref{hR^5-m5m})--
(\ref{hR^m-nrs}) we have $^{(5)}R^{\mu}_{\nu\,\rho\, z} = 0$ and
\begin{eqnarray}
&&^{(5)}R_{z\,\mu\, z\,\nu} = -
    2\epsilon V \left(2\zeta V_{\zeta\zeta}
    -{2\zeta\psi^{-1}\psi_{\zeta}V_{\zeta}}
    +{V_{\zeta}}\right)u_{\mu}^{\epsilon}u_{\nu}^{\epsilon}
    +2V\left({2\zeta\psi^{-1}\psi_{\zeta}V_{\zeta}}
  +{V_{\zeta}}\right)h_{\mu\nu}\,,\label{5R^5i5kz}\\
 && ~ \qquad  ^{(5)}R_{\mu\,\nu\,\rho\,\sigma}
  ={G}{ \zeta^{-2}}\,
  \left(h_{\mu\sigma}h_{\nu\rho}
   -h_{\mu\rho}h_{\nu\sigma}\right) + \nonumber \\
 && + 4\epsilon\psi\left(\zeta\psi_{\zeta\zeta}+\psi_{\zeta}
    -{\zeta\psi^{-1}\psi^{2}_{\zeta}}
   \right) \left(u_{\mu}^{\epsilon}(u^{\epsilon}_{\rho}h_{\nu\sigma}
   -u^{\epsilon}_{\sigma}h_{\nu\rho})
   - u^{\epsilon}_{\nu}(u^{\epsilon}_{\rho}h_{\mu\sigma}
   -u^{\epsilon}_{\sigma}h_{\mu\rho}) \right)\,.\label{5hR^m-nrsz}
\end{eqnarray}
Here $G =4\zeta^2\psi_{\zeta} \left(\psi+\zeta \psi_{\zeta}\right)
=\mbox{const}$. From here, using (\ref{5pszz}) and (\ref{5Vzz}), we obtained
\begin{eqnarray}
 & ^{(5)}R_{z\mu\, z\,\nu}  =
 -{2VV_{\zeta}}{\psi^{-1}}\left(\psi+2\zeta\psi_{\zeta}\right)
    (3\varepsilon u_{\mu}u_{\nu}+h_{\mu \nu})\,,\label{5R5i5k} \\
 & ^{(5)}\!R_{\mu\,\nu\,\rho\,\sigma}=
   4\psi_{\zeta}(\psi+\zeta\psi_{\zeta})
    \left(h_{\mu\sigma}h_{\nu\rho}
   -h_{\mu\rho}h_{\nu\sigma} +\right. \qquad \\ \nonumber
 & \left. + \epsilon u_{\sigma}(u_{\mu}h_{\nu\rho}
   -u_{\nu}h_{\mu\rho}) - \epsilon u_{\rho}(u_{\mu}h_{\nu\sigma}
   -u_{\nu}h_{\mu\sigma}) \right)\,.\label{5Rlijk}
\end{eqnarray}
Using the conservation law $G =\mbox {const} $ and equations (\ref{5pszz}) and
(\ref{5Vzz}), we have
\begin{equation}\label{psiz}
   \psi_{\zeta} \left(\psi+\zeta
\psi_{\zeta}\right)={G}/{4\zeta^2}\,,\quad
   V_{\zeta}\,(\psi+2\zeta\psi_{\zeta}) =-{G
    V} /{2\psi\zeta^2}\,.\label{5Vza}
\end{equation}
Therefore, the components of the curvature tensor (\ref{5R5i5k}),
(\ref{5Rlijk}) can be rewritten as
\begin{eqnarray}
  & ^{(5)}R_{z\mu\,z\,\nu}  = {G V^2}\zeta^{-2}\psi^{-2}
    (3\varepsilon u_{\mu}u_{\nu}+h_{\mu \nu})\,,\label{5R5i5kG}\\
  & ^{(5)}\!R_{\mu\,\nu\,\rho\,\sigma}={G}{\zeta^{-2}}
    \left(h_{\mu\sigma}h_{\nu\rho}
   -h_{\mu\rho}h_{\nu\sigma}
  + \epsilon u_{\sigma}(u_{\mu}h_{\nu\rho}
   -u_{\nu}h_{\mu\rho}) - \epsilon u_{\rho}(u_{\mu}h_{\nu\sigma}
   -u_{\nu}h_{\mu\sigma}) \right)\,.\label{5RlijkG}
\end{eqnarray}
Hence we find an invariant of the curvature tensor
\begin{eqnarray}\label{5RimG}
  I= ^{(5)}\!R_{ABCD}{^{(5)}\!R}^{ABCD}
  = 4{^{(5)}R}_{z\mu\, z\,\nu}{^{(5)}R}^{z\mu\,z\,\nu}
  +^{(5)}\!R_{\mu\,\nu\,\rho\,\sigma}
  {^{(5)}\!R}^{\mu\,\nu\,\rho\,\sigma}
  =72G^2\zeta^{-4}\psi^{-8}\,
\end{eqnarray}
or, using the explicit form of the metric (\ref{dspm}), we obtain
\begin{equation}\label{5Rim2}
  I= {72G^2}\zeta^{-4} \left(1-{ G}/{4\zeta}\right)^{-8}
  = 72 G^2 \left(u_{\epsilon}-{\epsilon G}/{4u_{\epsilon}}\right)^{-8}\,.
\end{equation}

It is easy to see that $\lim_{\zeta\rightarrow 0}I=0\,,\ \lim_{\zeta\rightarrow
\pm \infty}I=0$. Thus, on a light cone and infinity the metric (\ref{dsue}) is
regular. In the case $\epsilon G<0$ the surfaces $|\zeta|=|u^2|=|G|/4$ are
surfaces of extreme curvature $ I_{\max}={72}/{G^2}$ and in the case $\epsilon
G>0$ they are surfaces of singular curvature $\lim_{|u^2|\rightarrow \epsilon
G/4} I=\infty $.

\section{Five-dimensional version of the Birkhoff theorem}\label{appen4}

In terms of differential geometry the metric tensor of manifold $V^{(2)}\subset
V^{(5)}$: $y^i=\mbox{const}$, according to (\ref{2metrb}), has the form
\begin{equation}
   g=\frac{\psi^{2}(z,\zeta)}{4\zeta}\, d\zeta\otimes d\zeta
    -V^{2}(z,\zeta)dz\otimes dz\,. \label{g1a3}
\end{equation}
The metric tensor determines the scalar product of vectors $\vec{A},\vec{B}$
and one-forms $\alpha,\beta$: $
  (\vec{A}\,,\vec{B})=g(\vec{A}\,,\vec{B})\,, \
  \langle\alpha\,,\beta\rangle=g^{-1}(\alpha\,,\beta)$.
Here
\begin{equation}\label{g1a3inv}
  g^{-1}=\frac{4\zeta}{\psi^{2}}\,
  \partial_{\zeta}\otimes \partial_{\zeta}
    -\frac{1}{V^{2}}\partial_{z}\otimes \partial_{z}\,.
\end{equation}

Let us introduce a new basis of the differential forms
$\{\omega_{L}\,,\omega_{\xi}\}$ as follows
\begin{equation}\label{om-xi}
  \omega_{L} =i_{\vec{\xi}}\,\Omega =-\Omega(.,\vec{\xi})\,,
  \qquad \omega_{\xi}=g(.,\vec{\xi})
\end{equation}
where $\Omega =(1/2){\psi V}(|\zeta|)^{-1/2}\,d\zeta\wedge d z $ is a
two-form of volume on $V^{(2)}$ and $\vec{\xi}$ is the Killing vector
(\ref{kilfin1}). We have
\begin{eqnarray}
  \omega_{L}=-\epsilon(L_{,\,\zeta}d\zeta+L_{,\,z}d z)
  = -\epsilon d L \,, \label{omkil1} \\
  \omega_{\xi} =-\frac{\psi^2 L_{,\,z}\,d\zeta
  +4\zeta V^2 L_{,\,\zeta}\,d z}
  {2\psi V\sqrt{\epsilon\zeta}}\,.\label{omkil2}
\end{eqnarray}
According to the Frobenius theorem \cite{kram}, \cite{frolov}, we
have $\omega_{\xi} =-\beta d \acute{z}$ where $\acute{z}$ and $\beta$
are some functions $\{z\,,\zeta\}$. Using the variables $\acute{z}$
and $L$ as new coordinates, we obtain
 $ \beta \gamma^{L\acute{z}}=\beta \langle d L,d\acute{z}\rangle
  =\epsilon\langle\omega_{L},\omega_{\xi}\rangle=0 $.
Then we find
\begin{equation}\label{scalprodxx}
  (\vec{\xi})^2=g(\vec{\xi}\,,\vec{\xi})=-\epsilon(\nabla
  L)^2=-\epsilon \gamma^{LL}\,.
\end{equation}
On the other hand,
\begin{equation}\label{scalprodxx1}
  (\vec{\xi}\,)^2=g(\vec{\xi}\,,\vec{\xi})
  =\langle \omega_{\xi},\omega_{\xi}\rangle
  =\beta^2 \gamma^{\acute{z}\acute{z}}\,.
\end{equation}
Hence we obtain $ \gamma^{LL}=-\epsilon\beta^2
\gamma^{\acute{z}\acute{z}}$. Therefore, we have
$\gamma_{\acute{z}\acute{z}}=- \epsilon\beta^2 \gamma_{LL}$. In
coordinates $\{\acute{z}\,, L\}$ the metric tensor (\ref{g1a3}) of
space $V^{(2)}$ becomes
\begin{eqnarray}\label{2ds3a}
  g = \gamma_{LL}( d L\otimes d L-\epsilon\beta^2
  d\acute{z}\otimes d\acute{z})\,,\quad
   g^{-1} =\gamma^{-1}_{LL}
   (\partial_{L}\otimes \partial_{L}- \frac{\epsilon}{\beta^2}
  \partial_{\acute{z}}\otimes \partial_{\acute{z}}) \,.
\end{eqnarray}
Then for the Killing vector (\ref{kilfin1})  we find $
\vec{\xi}=g^{-1}(.,\omega_{\xi})
={\epsilon}{\beta^{-1}\gamma^{-1}_{LL}}\,\partial_{\acute{z}}$. From
$(L,L)$-component of the Killing equations for the metric
(\ref{2ds3a}) it follows $\gamma_{LL}=\gamma_{LL}(L)$. Note that
$(z,z)$-component of the Killing equations are satisfied identically.
From $(z,L)$-component of the Killing equations the relation
${\partial (\beta\gamma_{LL})}/{\partial {L}} =0$ is obtained. Hence
$\beta=f(\acute{z})/\gamma_{LL}$ where $f(\acute{z})$ is a certain
function of $\acute{z}$. In addition, we can suppose that
$f(\acute{z})=1$. Thus, in the coordinates $\{\acute{z},\,L\}$ the
metric (\ref{pmmetr1}) takes the form
\begin{equation}\label{5ds3a}
  ^{(5)}ds^2 =- \epsilon \gamma^{-1}_{LL}\,d\acute{z}^2+
  \gamma_{LL}(L)dL^2
   - L^2 \,{^{(3)}\!d\Omega}^2_{\epsilon}\,.
\end{equation}
The unknown function $\gamma_{LL}$ can be found from the conservation
law (\ref{5mass}):
\begin{equation}\label{solgll3a}
    \gamma^{LL}=
    \gamma^{ab} L_{,\, a} L_{,\, b}=(\nabla L)^2
    =\epsilon+G L^{-2}\,.
\end{equation}
Finally the metric (\ref{5ds3a}) can be written as \cite{gladush7}
\begin{equation}\label{pmmetrfin}
  ^{(5)}ds^2 =
  -\epsilon\left(\epsilon+{G}/{L^2}\right)\,d\acute{z}^2+
  \left(\epsilon+{G}/{L^2}\right)^{-1}dL^2
  - L^2 \,{^{(3)}\!d\Omega}^2_{\epsilon}\,.
\end{equation}

Let us now return from the coordinates $\{\acute{z},\,L\}$ to the conformal
coordinates $\{z,\,u_{\epsilon}\}$ in which the space-time part of the metric
takes the conformally flat form (\ref{pmmetr}). As a result of substituting $L=
L(u)$ into (\ref{5ds3a}), we get
\begin{equation}\label{soltrans3a}
  ^{(5)}ds^2 =- \frac{\epsilon}{\gamma_{LL}}\,d\acute{z}^2+
  \gamma_{LL}\left(\frac{d L}
  {d u_{\epsilon}}\right)^2du_{\epsilon}^2
  -{L^2}\,^{(3)}d\Omega^{2}_{\epsilon}\,.
\end{equation}
Comparing this metric with the metric (\ref{pmmetr}), we conclude that
\begin{equation}\label{compare}
  \gamma_{LL}L_{u_{\epsilon}}^2=\epsilon \psi^2\,, \qquad
     L=u_{\epsilon}\psi\,, \qquad
    V^2={\epsilon}/{\gamma_{LL}}\,.
\end{equation}
From here it follows that $
  \pm{u_{\epsilon}}\sqrt{\epsilon\gamma_{LL}} {d L}
  ={L}{du_{\epsilon}}$.
In the case $\epsilon\gamma_{LL}>0$ it has the a solution in the implicit form
\begin{equation}\label{solLu3a}
u_{\epsilon}=C\exp\int\,d L\left(\pm\,{\sqrt{\epsilon\gamma_{LL}}}/
  {L}\right)\,.
\end{equation}
Thus, from the metric (\ref{5ds3a}) we can return to the metric
(\ref{pmmetr}) wherein all functions do not depend on a fifth
coordinate $z $ this proves the statement.



\begin{thebibliography}{99}

\def\CMPh{{ Commun. Math. Phys.}~}
\def\JPh{{ J. Phys.}~}
\def\CJP{{ Czech. J. Phys.}~}
\def\LMPh {{ Lett. Math. Phys.}~}
\def\NPh  {{ Nucl. Phys.}~}
\def\PhE  {{ Phys. Essays}~}
\def\PhL  {{ Phys. Lett.}~}
\def\PhR  {{ Phys. Rev.}~}
\def\PhRL {{ Phys. Rev. Lett.}~}
\def\PhRp {{ Phys. Rep.}~}
\def\NCim {{ Nuovo Cimento}~}
\def\NuPB {{ Nucl. Phys.}~}
\def\GRG {{ Gen. Relativ. Gravit.}~}
\def\CQG {{ Class. Quantum Grav.}~}
\def\prp {report}
\def\Prp {Report}
\def\GrC {{Gravitation$\&$Cosmology}~}
\def\DANS {{Dokl. Akad. Nauk SSSR}~}
\def\APh {{Ann. Phys.}~}
\def\JMM {{Journ. Math. and Mech.}~}
\def\JMP {{J. Math. Phys.}~}
\def\IVUZ {{\it Izv. Vyssh. Uchebn. Zaved. Fiz}~}
\def\APP {{ Acta Phys. Pol.}~}
\def\UFZh {{  Ukr. Fiz. Zh.}~}
\def\TMF {{ Teor. Mat. Fiz.}~}
\def\ZEMF {{ Zh. Eksp. Teor. Fiz.}~}
\def\SPhJ {{ Sov. Phys. J.}~}
\def\SPhJETP {{ Sov. Phys. JETP.}~}
\def\PhEs {{ Phys. Essays}~}
\def\RMP {{Rev. Mod. Phys.}~}
\def\IJMD {{Int. Journ. Mod. Phys.}~}

\bibitem{turner}
M.~S.~Turner, Astron. Soc. Pac. Conf. Series. {\bf 1}, 666 (1999).
\bibitem{weinberg}
S.~Weinberg, \RMP {\bf 61}, 1 (1989).
\bibitem{chernin}
A.~D.~Chernin, Usp. Fiz. Nauk {\bf 171}, 1153 (2001).
\bibitem{sahni}
V.~Sahni and A.~Starobinsky, \IJMD {\bf D9}, 373 (2000).
\bibitem{zeldovich}
Ya.~B.~Zeldovich, {Usp. Fiz. Nauk}. {\bf 95}, 209 (1968).
\bibitem{gliner1}
E.~B.~Gliner, \ZEMF {\bf 49}, 542 (1965);\\ \DANS {\bf 192}, 771
(1970).
\bibitem{gliner2}
E.~B.~Gliner and I.~G.~Dymnikova, { Lett. Astr. J.} {\bf 1}, 7
(1975).
\bibitem{guth}
A.~H.~Guth, \PhR {\bf D23}, 347 (1981).
\bibitem{linde90}
A.~D.~Linde, {\it Particle physics and inflationary cosmology},
(Harwood academic press, Switzerland, 1990) e-Print Archive:
hep-th/9410082.
\bibitem{brandenberger}
R.~H.~Brandenberger, {\it Inflationary cosmology: progress and
problems}, (Proceedings International School on Cosmology, Kish
Island, Iran (Kluwer, Dordrecht) 2000)\\ e-Print Archive:
hep-ph/9910410.
\bibitem{hawking}
S.~W.~Hawking and G.~F.~R.~Ellis, {\it The Large Scale Structure of
Space-Time} (Cambridge University Press, Cambridge, 1973).
\bibitem{gladush7}
V.~D.~Gladush, \TMF {\bf 136}, 480 (2003) [Theor.Math.Phys. {\bf
136}, 1312 (2003)].
\bibitem{karl}
B.~Karliga, {Contributions to Algebra and Geometry}, {\bf 37}, 329
(1996).
\bibitem{petrov}
A.~Z.~Petrov, {\it Einstein spaces}, (Pergamon, Oxford, 1969).
\bibitem{wes}
J.~M.~Overduin and P.~S.~Wesson, \PhRp {\bf 283}, 303 (1997).
\bibitem{kram}
D.~Kramer, H.~Stephani, M.~Maccallum and E.~Herlt; Ed. E.Schmutzer,
{\it Exact Solution of the Einstein's Field Equations}, (Cambridge,
Cambridge University Press, 1980).
\bibitem{frolov}
V.~P.~Frolov and I.~D.~Novikov, {\it Black Hole Pysics: Basic
Concepts and New Developments}, (Dordrecht, Netherlands: Kluwer
Academic, 
1998).
\bibitem{cavag}
M.~Cavaglia, V.~Alfaro, and A.~T.~Filippov, Phys.Lett. {\bf B424},
265 (1998);
\\  M.~Cavaglia,
\PhR {\bf D59}, 084011 (1999); 
\\ D.~Grumiller, W.~Kummer, and D.~V.~Vassilevich,
\PhRp {\bf 369}, 327 (2002).
\bibitem{misner}
C.~W.~Misner, and D.~H.~Sharp,  \PhR {\bf B136}, 571 (1964).
\bibitem{hernandes}
W.~C.~Hernandes and G.~C.~Misner, {Astrph. J.} {\bf 143}, 452 (1968);
\\M.~E.~Cahill and G.~C.~McVittie, \JMP {\bf 11}, 1382 (1970).
\bibitem{matrtinez}
E.~A.~Martinez and Jr.~J.~W.~York, \PhR {\bf D 40}, 2124 (1989).
\bibitem{gladush1}
V.~D.~Gladush, {Acta Phys. Pol.} {\bf B30}, 3 (1999);\\ V.~D.~Gladush
and R.~A.~Konoplya, {J. Math. Phys.} {\bf 40}, 955 (1999).
\bibitem{gladush2}
V.~D.~Gladush, Izv. Vyssh. Uchebn. Zaved. Fiz. {\bf N11}, 58 (1979)
\\  {[English translation: Sov. Phys. J. {\bf 22}, 1172 (1979)]}.
\bibitem{gang}
A.~Gangui, {\it Topological Defects in Cosmology}, (Lecture Notes for
the First Bolivian School on Cosmology (2001)) e-Print Archive:
astro-ph/0110285.
\bibitem{fronsdal}
C.~Fronsdal, \PhR  {\bf 116}, 778 (1959);
\\ J.~Rosen, \RMP {\bf 37}, 204 (1965).
\bibitem{david}
A.~Davidson,  \PhR {\bf D2}, 1811 (1987).
\bibitem{haar}
D.~Ter Haar,  {\it Elements of Hamiltonian mechanics} (Pergamon,
Oxford, 1971).

\end{thebibliography}
\end{document}